\documentclass[trackchanges,twocolumn]{aastex7}

\newcommand{\Sg}{Sgr~A*}
\defcitealias{vonFellenberg2025}{Paper~I}
\defcitealias{vonFellenberg2025_extinction}{MIR Extinction Paper}
\usepackage{amsmath, amsfonts, amssymb}
\usepackage{gensymb}

\begin{document}

\title{First Mid-infrared Detection and Modeling of a Flare from Sgr A*. II. Mid-IR Spectral Energy Distribution and Millimeter Polarimetry}

\author[orcid=0000-0003-3503-3446,sname=Michail, gname=Joseph]{Joseph M. Michail}
\altaffiliation{NSF Astronomy \& Astrophysics Postdoctoral Fellow}
\affiliation{Center for Astrophysics $|$ Harvard \& Smithsonian, 60 Garden Street, Cambridge, MA 02138, USA}
\email[show]{joseph.michail@cfa.harvard.edu}  

\author[orcid=0000-0000-0000-0001,sname='von Fellenberg', gname=Sebastiano]{Sebastiano D. von Fellenberg}
\altaffiliation{Feodor Lynen Fellow}
\affiliation{Canadian Institute for Theoretical Astrophysics, University of Toronto, 60 St.\ George Street, Toronto, ON M5S 3H8, Canada}
\affiliation{Max Planck Institute for Radioastronomy, auf dem H{\"u}gel 69, Bonn, Germany }
\email[show]{sfellenberg@cita.utoronto.ca}  

\author[orcid=0000-0002-3490-146X,sname=Keating, gname=Garrett]{Garrett K. Keating}
\affiliation{Center for Astrophysics $|$ Harvard \& Smithsonian, 60 Garden Street, Cambridge, MA 02138, USA}
\email{garrett.keating@cfa.harvard.edu}

\author[orcid=0000-0002-1407-7944,sname=Rao, gname=Ramprasad]{Ramprasad Rao}
\affiliation{Center for Astrophysics $|$ Harvard \& Smithsonian, 60 Garden Street, Cambridge, MA 02138, USA}
\email{rrao@cfa.harvard.edu}

\author[0009-0003-9906-2745]{Tamojeet Roychowdhury}
\affiliation{Department of Astronomy, University of California Berkeley, Berkeley, CA 94704, USA}
\email{tamojeet@berkeley.edu}

\author[0000-0002-9895-5758]{S. P. Willner}
\affiliation{Center for Astrophysics $|$ Harvard \& Smithsonian, 60 Garden Street, Cambridge, MA 02138, USA}
\email{swillner@cfa.harvard.edu}

\author[0000-0001-8921-3624]{Nicole M. Ford}
\affiliation{McGill University, Montreal QC H3A 0G4, Canada}%
\affiliation{Trottier Space Institute, 3550 Rue University, Montréal, Québec, H3A 2A7, Canada}
\email{nicole.ford@mail.mcgill.ca}

\author[0000-0001-6803-2138]{Daryl Haggard}
\affiliation{McGill University, Montreal QC H3A 0G4, Canada}%
\affiliation{Trottier Space Institute, 3550 Rue University, Montréal, Québec, H3A 2A7, Canada}
\email{daryl.haggard@mcgill.ca}

\author[0000-0001-9564-0876]{Sera Markoff}
\affiliation{Anton Pannekoek Institute for Astronomy, University of Amsterdam, Science Park 904, 1098 XH Amsterdam, The Netherlands}
\affiliation{Gravitation and Astroparticle Physics Amsterdam Institute, University of Amsterdam, Science Park 904, 1098 XH 195 196 Amsterdam, The Netherlands}
\email{S.B.Markoff@uva.nl}

\author[0000-0001-7801-0362]{Alexander Philippov}
\affiliation{University of Maryland, College Park, MD 20742, USA}
\email{sashaph@umd.edu}

\author[0000-0002-7301-3908]{Bart Ripperda}
\affiliation{Canadian Institute for Theoretical Astrophysics, University of Toronto, 60 St.\ George Street, Toronto, ON M5S 3H8, Canada}
\affiliation{Dunlap Institute for Astronomy and Astrophysics, University of Toronto, 50 St.\ George Street, Toronto, ON M5S 3H4, Canada}
\affiliation{Department of Physics, University of Toronto, 60 St. George Street, Toronto, ON M5S 1A7, Canada.}
\email{bartripperda@gmail.com}

\author[]{Sophia S\'anchez-Maes}
\affiliation{University of Maryland, College Park, MD 20742, USA}
\email{sophiasm@umd.edu}

\author[0009-0004-8539-3516]{Zach Sumners}
\affiliation{McGill University, Montreal QC H3A 0G4, Canada}%
\affiliation{Trottier Space Institute, 3550 Rue University, Montréal, Québec, H3A 2A7, Canada}
\email{ronald.sumners@mail.mcgill.ca}

\author[0000-0003-2618-797X]{Gunther Witzel}
\affiliation{Max Planck Institute for Radioastronomy, auf dem H{\"u}gel 69, Bonn, Germany }
\email{gwitzel@mpifr-bonn.mpg.de}

\author[0000-0001-9641-6550]{Mayura Balakrishnan}
\affiliation{McGill University, Montreal QC H3A 0G4, Canada}%
\affiliation{Trottier Space Institute, 3550 Rue University, Montréal, Québec, H3A 2A7, Canada}
\email{mayura.balakrishnan@mcgill.ca}

\author[0000-0002-8776-1835]{Sunil Chandra}
\affiliation{Physical Research Laboratory, Mt Abu Observatory, Rajasthan, 307501, India}
\email{sunil.chandra355@gmail.com}

\author[0000-0001-6906-772X]{Kazuhiro Hada}
\affiliation{Graduate School of Science, Nagoya City University,  Yamanohata 1, Mizuho-cho, Mizuho-ku, Nagoya, 467-8501, Aichi, Japan}
\affiliation{Mizusawa VLBI Observatory, National Astronomical Observatory of Japan, 2-12 Hoshigaoka, Mizusawa, Oshu, Iwate 023-0861, Japan}
\email{hada@nsc.nagoya-cu.ac.jp}

\author[0000-0003-4801-0489]{Macarena Garcia Marin}
\affiliation{European Space Agency (ESA), ESA Office, Space Telescope Science Institute, 3700 San Martin Drive, Baltimore, MD 21218, USA}
\email{Macarena.Garcia.Marin@esa.int}

\author[0000-0003-0685-3621]{Mark A. Gurwell}
\affiliation{Center for Astrophysics $|$ Harvard \& Smithsonian, 60 Garden Street, Cambridge, MA 02138, USA}
\email{mgurwell@cfa.harvard.edu}

\author[0000-0002-0670-0708]{Giovanni G. Fazio}
\affiliation{Center for Astrophysics $|$ Harvard \& Smithsonian, 60 Garden Street, Cambridge, MA 02138, USA}
\email{gfazio@cfa.harvard.edu}

\author[0000-0002-5599-4650]{Joseph L. Hora}
\affiliation{Center for Astrophysics $|$ Harvard \& Smithsonian, 60 Garden Street, Cambridge, MA 02138, USA}
\email{jhora@cfa.harvard.edu}

\author[orcid=0009-0001-1040-4784]{Braden Seefeldt-Gail}
\affiliation{Canadian Institute for Theoretical Astrophysics, University of Toronto, 60 St.\ George Street, Toronto, ON M5S 3H8, Canada}
\email{braden.gail@mail.utoronto.ca}
\affiliation{Dunlap Institute for Astronomy and Astrophysics, University of Toronto, 50 St.\ George Street, Toronto, ON M5S 3H4, Canada}

\author[]{Howard A. Smith}
\affiliation{Center for Astrophysics $|$ Harvard \& Smithsonian, 60 Garden Street, Cambridge, MA 02138, USA}
\email{hsmith@cfa.harvard.edu}


%
%
\begin{abstract}
\citet[][Paper~I]{vonFellenberg2025} reported the first mid-infrared detection of a flare from Sgr~A*. The JWST/MIRI/MRS observations were consistent with an orbiting hotspot undergoing electron injection with a spectrum that subsequently breaks from synchrotron cooling. However, mid-infrared extinction measurements appropriate for these data were not yet determined, and therefore the temporal evolution of the absolute spectral index remained unknown. This work applies new Galactic Center extinction measurements to the flare observations. The evolution of the spectral index after the peak is fully consistent with that reported in Paper~I with a maximum absolute mid-infrared spectral index $\alpha_{\rm{MIR}}=0.45\pm0.01_{\rm{stat}}\pm0.08_{\rm{sys}}$ during the second mid-infrared flare peak, matching the known near-infrared spectral index during bright states ($\alpha_{\rm{NIR}}\approx0.5$). There was a near-instantaneous change in the mid-infrared spectral index of $\Delta\alpha_{\rm{MIR}}=0.33\pm0.06_{\rm{stat}}\pm0.11_{\rm{sys}}$ at the flare onset. We propose this as a quantitative definition for this infrared flare's beginning, physically interpreted as the underlying electron distribution's transition into a hard power-law distribution. This paper also reports the SMA millimeter polarization during the flare, which shows a small, distorted, but overall clockwise-oriented Stokes Q--U loop during the third mid-infrared peak. Extrapolating the mid-infrared flux power law to the millimeter yields a variable flux consistent with the observed 220 GHz emission. These results, together with the Paper I modeling, plausibly suggest a single hotspot produced both the mid-infrared and millimeter variability during this event. However, additional flares are required to make a general statement about the millimeter and mid-infrared connection.
\end{abstract}


\section{Introduction}\label{sec:intro}
    The Galactic Center supermassive black hole \Sg\ exhibits daily bright flares observed in the infrared \citep[IR, e.g.,][]{Genzel2003} and X-ray \citep[e.g.,][]{Neilsen2013}. It is also continuously variable in the IR \citep[e.g.,][]{Do2009,GravityCollaboration2020flux}, (sub)millimeter \citep[(sub)mm, e.g.,][]{Dexter2014}, and radio regimes \citep[e.g.,][]{Bower2015}.
    While the temporal correlation of X-ray flares with near-IR flares is clear \citep[e.g.,][]{Eckart2012}, the temporal connection between the near-IR and (sub)mm regime is not established. 
    Several analyses have shown temporal correlation \citep[e.g.,][]{Michail2021,Witzel2021}, temporal anti-correlation \citep[e.g.,][]{Dodds-Eden2010,Michail2024}, and no (or unclear) correlation \citep[e.g.,][]{Fazio2018, Boyce2019, Boyce2021} in multiwavelength campaigns of this enigmatic source. In recent years, consensus is building around a ``hotspot'' picture of flares \citep[e.g.,][]{Genzel2003,Broderick2006_hotspot,Hamaus2009}, in which a region of the accretion flow becomes luminous due to non-thermal particle acceleration, either through turbulence, reconnection, or both \citep[][]{Markoff2001,Yuan2009, Dodds-Eden2010}. More recently, flux tubes emerging in magnetically-arrested disks \citep[MAD;][]{narayan2003}, produced by flux ejection through magnetic reconnection near the black hole event horizon, have been identified as regions where the non-thermal acceleration may take place \citep[][]{Dexter2020_flare,Porth2021,Ripperda2022,Zhdankin2023}. One important observational signature of such hotspots is their dynamical behavior, as any stable, luminous, and coherent structure will orbit the black hole, observationally imprinting orbital signatures in the measured flux, polarimetry, and astrometry \citep[e.g.,][]{GravityCollaboration2020_polariflares, GravityCollaboration2020_orbital}. Such (quasi-) periodic behavior has been suggested since the first observations of flares \citep[e.g.,][]{Genzel2003}, many of which show double-peaked structures separated by 20--40 minutes, corresponding to the expected orbital time scales \cite[e.g.,][]{Haggard2019}. However, this interpretation has been challenged by the absence of statistically significant periodicities in light-curve power spectra. Instead, the stochastic nature of light curves has been used to argue against an event-like interpretation of flares \citep[e.g.,][]{Do2009,Meyer2014,Witzel2018}. On the other hand, \cite{vonFellenberg2024} showed that the absence of significant periodicity can be explained if the event duration is similar to the orbital time scale and the orbit is at a low inclination, suppressing lensing effects. For instance, an analysis by \citet{Iwata2020} showed a possibly static $\sim$35-minute quasi-periodicity in long-duration ALMA observations of \Sg. Likewise, Spitzer observations analyzed by \citet{Michail2024} demonstrated a $\sim$35-minute transient periodicity in the light curves during a triple-peaked flaring episode. 
    
    \citet[][]{GRAVITYCollaboration2018_orbital, GRAVITY_collab2023_polariflares} detected loop-like astrometric signatures during four bright near-IR flares, which were interpreted as orbiting hotspots with orbital radii $R_{\rm{orb}}\approx7$--$10~R_g$ and viewed at low inclination angles ($\sim$10--40\degree). This was further corroborated by polarimetric measurements which showed clockwise loops in the Stokes $Q$--$U$ plane, matching the astrometric signature if the hotspots have a dominantly-poloidal magnetic-field component \citep{GravityCollaboration2020_polariflares}; this poloidal-dominant magnetic field configuration is also found in the emitting regions of MAD flux tubes \citep[][]{Porth2021,Ripperda2022}.
    
    \cite{Wielgus2022} reported similar clockwise Stokes $Q$--$U$ loops in $220~\mathrm{GHz}$ ALMA observations carried out during Event Horizon Telescope observations in 2017 \citep{eht_sgra_I, eht_sgra_II} approximately half an hour after a bright X-ray flare. The prominent $Q$--$U$ loop was interpreted as an orbital signature, where the authors derived orbital parameters consistent with those found in the near-IR with GRAVITY\null. Similar $Q$--$U$ loops had been reported much earlier in Submillimeter Array (SMA) observations \citep{Marrone2008}, albeit with lower signal-to-noise and a counterclockwise rotation. Adiabatic expansion of hotspots may also produce secondary polarization variations atop those driven by orbital motion \citep{Michail2023}.
    
    \citet[][\citetalias{vonFellenberg2025}]{vonFellenberg2025} reported the first detection of a mid-IR flare between $\sim$~5 and $\sim$~21~\micron. The mid-IR spectral index $\alpha_{\rm{MIR}}$ ($\alpha \equiv d\log\left(\nu F_\nu\right)/d\log\nu$, where $F_\nu$ is the flux density at frequency $\nu$) is a sensitive discriminator between dominant emission mechanisms \citep[see, e.g.,][]{Dodds-Eden2009, Witzel2021}. The flare showed significant spectral steepening, interpreted as the signature of synchrotron cooling. Modeling the mid-IR flare emission as a hotspot within the accretion flow undergoing orbit-induced Doppler boosting and synchrotron cooling gave a magnetic field strength $B\sim40$--70~{G}, consistent with 
    the synchrotron cooling. Contemporaneous SMA observations at $220~\mathrm{GHz}$ found a correlated increase in the flux of $\Delta F_{\mathrm{220~\rm{GHz}}}=0.3~\mathrm{Jy}$, delayed $\approx$10 minutes after the final mid-IR peak. Because the SMA observed only the final $\approx$30 minutes of the mid-IR flare, this places a lower limit of 10 minutes on the actual time lag. Paper~I's simple model can also recover the mm-mid-IR time delay without requiring--but also not precluding--adiabatic expansion in typical jet models \citep[e.g.,][]{YusefZadeh2006, Markoff2007, Falcke2009, Brinkerink2021, Chavez2024}. The same, cooled, non-thermal population of electrons could have been responsible for the mm flare if that electron population was large enough. However, the flare's mid-IR flux density, and therefore the electron population, was poorly constrained because of the lack of accurate mid-IR extinction and photometric aperture corrections. Instead, Paper~I used relative flux measurements to model the mid-IR light curves in anticipation of a more suitable extinction law for these mid-IR data. Our companion paper \citep[][henceforth referred to as the \citetalias{vonFellenberg2025_extinction}]{vonFellenberg2025_extinction} reports a new mid-IR extinction law, allowing us to obtain a predictive and intrinsic flux calibration on an absolute scale for the mid-IR flare, overcoming this limitation.
    
    This paper describes the mid-IR, time-variable, absolute spectral energy distribution of the 2024 April 6 flare from Sgr A*. Additionally, we report and analyze the simultaneous millimeter polarimetric light curve taken with the SMA\null. Throughout this work, we assume the mass of \Sg\ is $M=4.152\times10^{6}~M_\odot$ (solar masses) at a distance of $8.178$ kpc \citep{GRAVITYCollaboration2019_geometricDistance}. This mass corresponds to a gravitational radius $R_g = GM/c^2=6.13\times10^{11}$ cm, where $G$ is the gravitational constant and $c$ is the speed of light. 

\section{Data and Calibration}\label{sec:data_cal}
    \subsection{Submillimeter Array (220 GHz)}\label{ssec:sma}
        The SMA observed \Sg\ for 4.2 hours on 2024 April 6 (UT) as part of the first multiwavelength campaign with JWST/MIRI (2023B-S017; PI: H. Smith). The SMA was in its extended configuration with baselines ranging from approximately 25 to 226 meters (18--166~$\text{k}\lambda$ at 220 GHz). The central frequency was 220.1 GHz with two basebands centered 10~GHz above (upper sideband, USB; central frequency of 230.1 GHz) and below (lower sideband, LSB; central frequency of 210.1 GHz) the central frequency. Each baseband was composed of six 2.288-GHz-wide spectral windows (SPWs) comprising 16384 channels recording full polarization products. We averaged (``rechunked'') the number of channels by a factor of 128, reducing the native channel resolution from 140 kHz to 17.875 MHz.
        
        \citetalias{vonFellenberg2025} published the total intensity SMA light curves, which were calibrated with standard interferometric tasks in CASA \citep{CASA}\null. This new analysis uses the SMA COMPASS pipeline (G. Keating et al., in prep.), which produces CASA-readable gain solutions and MeasurementSet (MS) files as required for calibrating the linear polarization products. SMA has a unique observing strategy by which the cross-hand bandpass and cross-receiver phases must be calibrated before instrumental polarization solutions can be produced. There is no set of standard CASA procedures to calibrate these terms outside of COMPASS, and the original reduction path required separate, but serial, calibration methods in separate calibration packages (``SMA-MIR'' and ``MIRIAD''). 
        
        MWC~349A and Ceres were observed as flux calibrators, and 3C~279 was used as the same- and cross-hand bandpass calibrator as well as the instrumental polarization (D) calibrator. We employed J1733$-$1304 (J1733) as the complex gain calibrator throughout the entire track. J1751+0939 (J1751) and J1924$-$2914 (J1924) were observed during the first and second half of the track, respectively, as secondary check sources. D-terms were derived with \texttt{polcal} using 3C~279 in CASA (version 6.4.1), then the complex gain tables, cross-receiver phase, and D-terms were applied using \texttt{applycal}. The data were split and phase self-calibrated with a point-source model located at the phase center per-integration, further removing atmospheric gain corruptions  that remain in the standard calibration, then exported to UVFITS for light-curve extraction in AIPS \citep{AIPS2003}. We extracted Stokes $I$, $Q$, and $U$ light curves using task \texttt{DFTPL} on projected baselines $\geq$30~k$\lambda$ at 60-second binning in the LSB, USB, and across the entire bandwidth. We derived the electric vector polarization angle (EVPA, $\chi$), de-biased linear-polarization fraction ($p_l$), and the spectral index from the data. The light curves, the associated flux and polarimetric quantities, and their mathematical definitions are presented in Appendix~\ref{ssec:sma_data_conv}. Additionally, given this new reduction pathway in CASA, we completed consistency checks between COMPASS-calibrated and AMAPOLA\footnote{\url{https://www.alma.cl/~skameno/AMAPOLA/}} results as described in Appendix~\ref{ssec:average_pol_results}.
        
    \subsection{JWST MIRI/MRS (5--21$~\mu$m)}\label{ssec:jwst}
        \subsubsection{Pipeline Calibration}
            JWST/MIRI observed \Sg\  on the same date during two $\sim$2-hour tracks coordinated with other multiwavelength facilities (PI: D. Haggard; PID 4572; \citealt{2023jwst.prop.4572H}) using the four \texttt{SHORT} gratings available to the Medium Resolution Spectrograph (MRS; \citealt{Wells2015}). This analysis focuses on the second track, originally published in \citetalias{vonFellenberg2025}\null. The light curves derived in that work used normalized, single-pixel photometry with normalization to remove the temporally-constant, multiplicative extinction and aperture corrections from the modeling. The present work uses the newly-derived extinction measurements from the \citetalias{vonFellenberg2025_extinction} and derives stare-based aperture corrections (described in Appendix \ref{appx:apcorr}) to determine the absolute, de-reddened photometric light curves of \Sg. 
            
    The MIRI/MRS \texttt{SHORT} gratings provide simultaneous spectral coverage in four mid-IR bands (channel 1: 4.90--5.74~\micron; channel 2: 7.51--8.77~\micron; channel 3: 11.55--13.47~\micron; channel 4: 17.70--20.95~\micron). 
    \citetalias{vonFellenberg2025} gives specific details of the non-standard calibration procedure required to extract per-integration light curves from MIRI/MRS\null.
    We reprocessed the data through the standard Level 2 and 3 JWST pipeline steps (version 1.17; \citealt{jwst_pipeline_bushouse2024}) with calibration context 1322, which provides spectral (S3D) cubes per integration. We developed a custom pipeline to ``rebuild'' these data into a time-sorted four-dimensional cube for further analysis. Our custom pipeline is built to fully process MRS light curves through these initial steps and can flexibly start with Level 2 (RATEINT) or 3 (CALINT) data files and is publicly available on 
    \href{https://github.com/JoeMichail/MRS_LightCurve_Calibration_Pipeline}{GitHub}.

        \subsubsection{Light Curve Extraction and Corrections}
            We adopted a two-step approach to extract light curves from \Sg, maximizing signal-to-noise while simultaneously ensuring minimal aperture corrections need to be applied. Smaller apertures centered on the most variable pixels are less susceptible to noise spikes than those derived with larger apertures. However, larger apertures ameliorate undersampling of the PSF \citep[due to the non-dithered nature of our observations; e.g.,][]{Law2025}, requiring smaller aperture corrections to obtain the intrinsic flux of \Sg. Following this, we chose two apertures (one small and one large) in each channel. The small apertures are square and encompass the most variable \Sg\  pixels in each of the channels; in channels 1 and 2, we adopt a $3\times3$ pix aperture ($0\farcs39\times0\farcs39$ and $0\farcs51\times0\farcs51$, respectively) and a $2\times2$ aperture in channels 3 and 4 ($0\farcs40\times0\farcs40$ and $0\farcs70\times0\farcs70$, respectively). Each of these small apertures contains the single pixel used by \citetalias{vonFellenberg2025} for the original light curves.
    
            For the larger apertures, we used a circular footprint centered on \Sg\  in each channel during the flaring event. We determined \Sg's location by measuring the flare centroid during one of the peaks in the event (first peak in channels 1--3, third peak in channel 4) relative to ``quiescence.'' The choice of the larger apertures' sizes was limited by the proximity of other features (i.e., the Mini-spiral and IRS sources) to \Sg, especially in channel 4, where the distance to the Mini-spiral's inner edge is approximately one full width half maximum (FWHM)\null. In channels 1 through 4, we adopted aperture radii of $0\farcs6$, $0\farcs7$, $0\farcs9$, and $1\farcs1$, respectively. As a secondary precaution, we flagged all pixels in channel 4 with surface brightness $\geq 76410$ MJy sr$^{-1}$ (220 mJy pix$^{-1}$), removing any remaining Mini-spiral pixels from the larger aperture.
    
           We performed aperture photometry in each aperture using the \texttt{aperture\_photometry} function in the astropy \texttt{photutils} package \citep{astropy:2013, astropy:2018, astropy:2022}. We also passed the uncertainty arrays produced in the JWST pipeline to obtain the uncertainty on the aperture photometry. We simultaneously de-trended and zero-point subtracted each spectral slice's light curve by using a linear fit to two time ranges that we define as ``quiescence'' ($t\in\left[10.79, 11.12\right]~\lor[12.51, 12.75]$ hours barycentric UT)\null. While de-trending is required for all four channels, it is most important in channel~4. The light-curve trend most likely caused by drifts in JWST's pointing, estimated from the Fine Guidance Sensor (FGS) centroid solutions of the guide star as ${\sim}250~\mu$as. We measured the detrending root-mean-square (RMS) uncertainty by drawing 1000 realizations of the linear fit parameters from the fit's covariance matrix. After detrending, we scaled the mean flux in each channel's smaller-aperture light curve to the larger aperture's mean flux. This limited artifacts in the light curves while simultaneously accounting for missing flux outside of the smaller aperture. To fully account for the missing flux outside of this larger aperture \citep[and the reconstructed IFU cubes;][]{Law2025}, we estimated and applied wavelength-dependent aperture correction in each of the chosen apertures for non-dithered observations, described in Appendix~\ref{appx:apcorr}\null. Finally, we applied extinction corrections derived from JWST/MIRI \citepalias{vonFellenberg2025_extinction} and averaged along each channel's wavelength axis to provide absolute, dereddened light curves at 86-second cadence. 

\begin{figure*}[!ht]
        \centering
        \includegraphics[width=\linewidth]{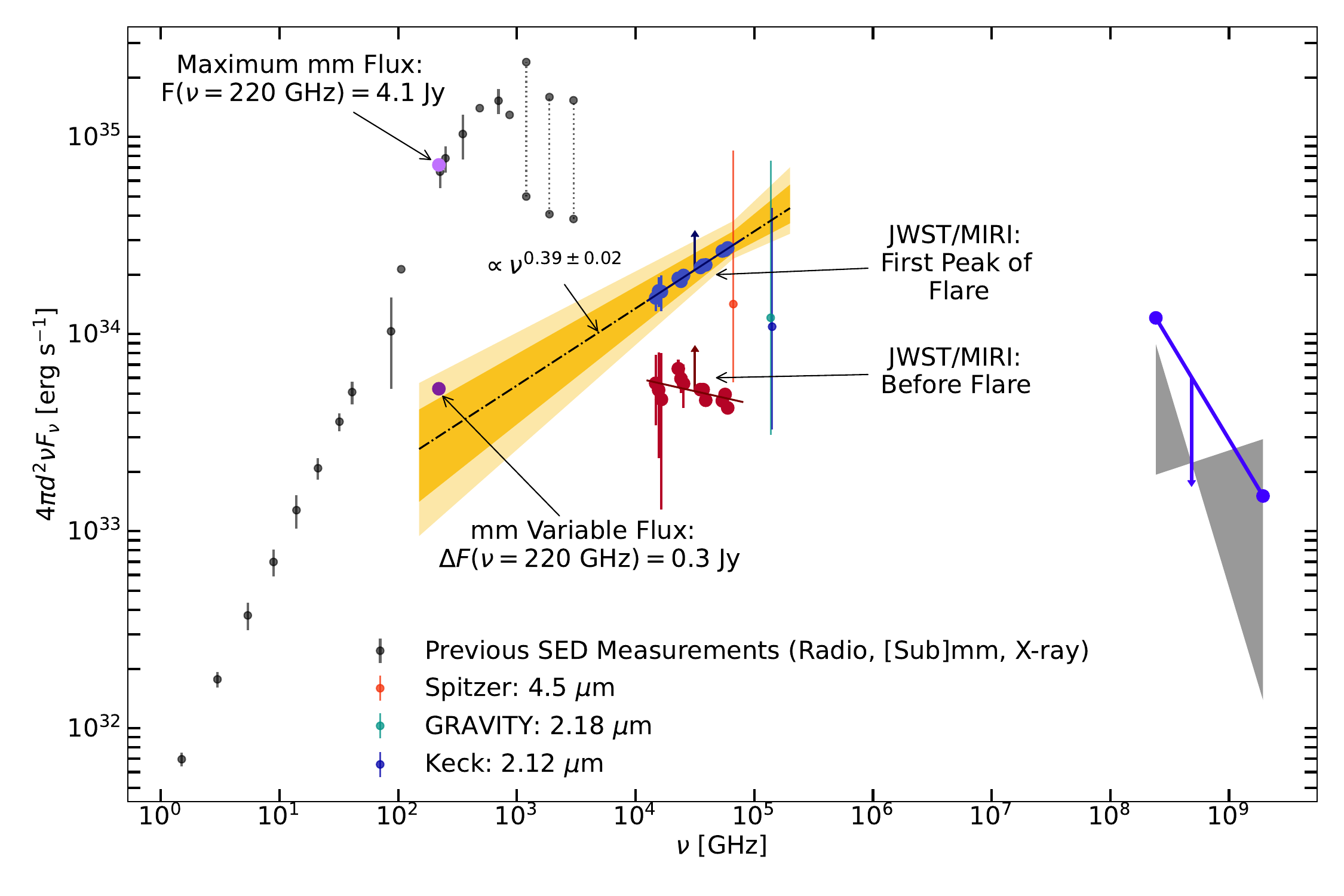}
        \caption{Radio to X-ray SED of \Sg, including mid-IR JWST/MIRI lower limits (red and blue points; corresponding times are denoted in Figure \ref{fig:timed_sed}), observed absolute maximum (light purple point) and variable (purple point) mm fluxes measured with the SMA, and X-ray flare upper limit from Chandra (blue-violet line) during this event. The first mid-IR peak's best-fit power law is extrapolated, and the 68\% (dark gold) and 90\% (light yellow) credible intervals are plotted. The variable mm data are fully within the 90\% credible range of the single power law stretching between mm to mid-IR. The binned, historic SED data are shown in the radio and mm \citep{Bower2015,Bower2019, liu_linearly_2016, Brinkerink2016}, far-IR \citep{Stone2016, VonFellenberg2018}, and X-ray \citep[quiescent ``bow-tie'' spectrum;][]{Baganoff2003_spectrum}. Spitzer \citep{Witzel2018}, GRAVITY \citep{GravityCollaboration2020flux}, and post-2019 Keck \citep{Weldon2023} points show the median flux and 5\% to 95\% percentile ranges. We have recalculated the ordinate using a standard distance of $d = 8.178$ kpc; for the Spitzer data, we have dereddened the data with the extrapolated extinction from the \citetalias[][$\rm{A}_{4.5~\mu\rm{m}} = 1.14$ mag]{vonFellenberg2025_extinction}.}
        \label{fig:mwl_sed}
\end{figure*}
    
\section{Mid-IR Spectral Energy Distribution}\label{sec:sed}
    \subsection{Broadband SED and Correlation to the Millimeter Variability}
        Figure \ref{fig:mwl_sed} shows the broadband spectral energy distribution (SED) of \Sg, including JWST/MIRI (at two different instances in the observation) and 220 GHz SMA\null. Historic SED values at radio, mm, far-IR, and X-ray are also presented. To more clearly show the mid-IR spectral index, we fit and plotted a power law to each of these two instances. In the radio through far-IR, the measured SEDs are a combination of variable and thermal ``background'' emission. The near- and mid-IR are predominantly produced by variable emission, and Figure~\ref{fig:mwl_sed} shows the $\Delta F_{220~\rm{GHz}}=300$ mJy variable component of the flare (dark purple) as well as the much larger total mm emission. 
        
        The ``before flare'' spectrum is centered at 11:27:20 UT with spectral index $\alpha_{\rm{MIR}}=-0.14\pm0.07_{\rm{stat}}\pm0.08_{\rm{sys}}$. The spectrum of the flare's first peak at 11:38:49 UT has a measured spectral index $\alpha_{\rm{MIR}}=0.39\pm0.01_{\rm{stat}}\pm0.08_{\rm{sys}}$. The times corresponding to these two spectra are marked in Figure \ref{fig:timed_sed} below. The statistical uncertainty on $\alpha_{\rm{MIR}}$ came from quadrature summing the unweighted statistical fit (from the fitting covariance matrix) and the standard error of the mean $\alpha_{\rm{MIR}}$ from 1000 flux-resampling Monte Carlo draws. The systematic uncertainty was calculated as the standard deviation of the derived $\alpha_{\rm{MIR}}$ values from each of the posterior extinction curves \citepalias{vonFellenberg2025_extinction}. The ``total'' error bar accounting for statistical + systematic errors is the quadrature sum of both error estimates.
        
        For mid-IR spectra plotted in this work, the presented values are lower limits on the true luminosity as the detrending removes an arbitrary baseline that likely contains non-flaring emission from \Sg. Figure \ref{fig:mwl_sed} also plots the median and 90\% credible interval ranges for Spitzer \citep{Witzel2018}, GRAVITY \citep{GravityCollaboration2020flux}, and Keck \citep[post-2019 values;][]{Weldon2023}. For more direct comparison with these other observatories, Figure~\ref{fig:mwl_sed} shows the first peak's spectrum extrapolated to the mm and to the near-IR. 

        Our Monte Carlo simulations predict a 68\% credible interval of $\Delta F_{\rm{220~GHz}}=165^{+99}_{-68}$ mJy at 220 GHz from the first mid-IR peak, which, at the upper bound, broadly agrees with the observed $\Delta F_{\rm{220~GHz}}=300$ mJy change. Expanding this range to 90\% credible interval ($\Delta F_{\rm{220~GHz}}=67$--351~mJy) includes the flux of the mm variability. This estimate corroborates the finding from \citetalias{vonFellenberg2025}, which found the extrapolated mm flux to be compatible with the normalized, mid-IR-only fit when the electron spectrum's low-energy cutoff is set to $\gamma_{\rm{min}}=10$. (Otherwise no appreciable mm flux is produced.) While this extrapolation is a sufficient approximation for this estimate, it does not account for electron cooling and Doppler boosting, which affect and increase the observed mm variability \citepalias{vonFellenberg2025}; a dynamic model, which is planned for future work, is required to account for time-dependent physical mechanisms such as electron injection and synchrotron cooling.
                
    \subsection{Spectral index evolution \& spectral-index-based definition of flares}
        \begin{figure*}[t]
            \centering
            \includegraphics[width=\linewidth]{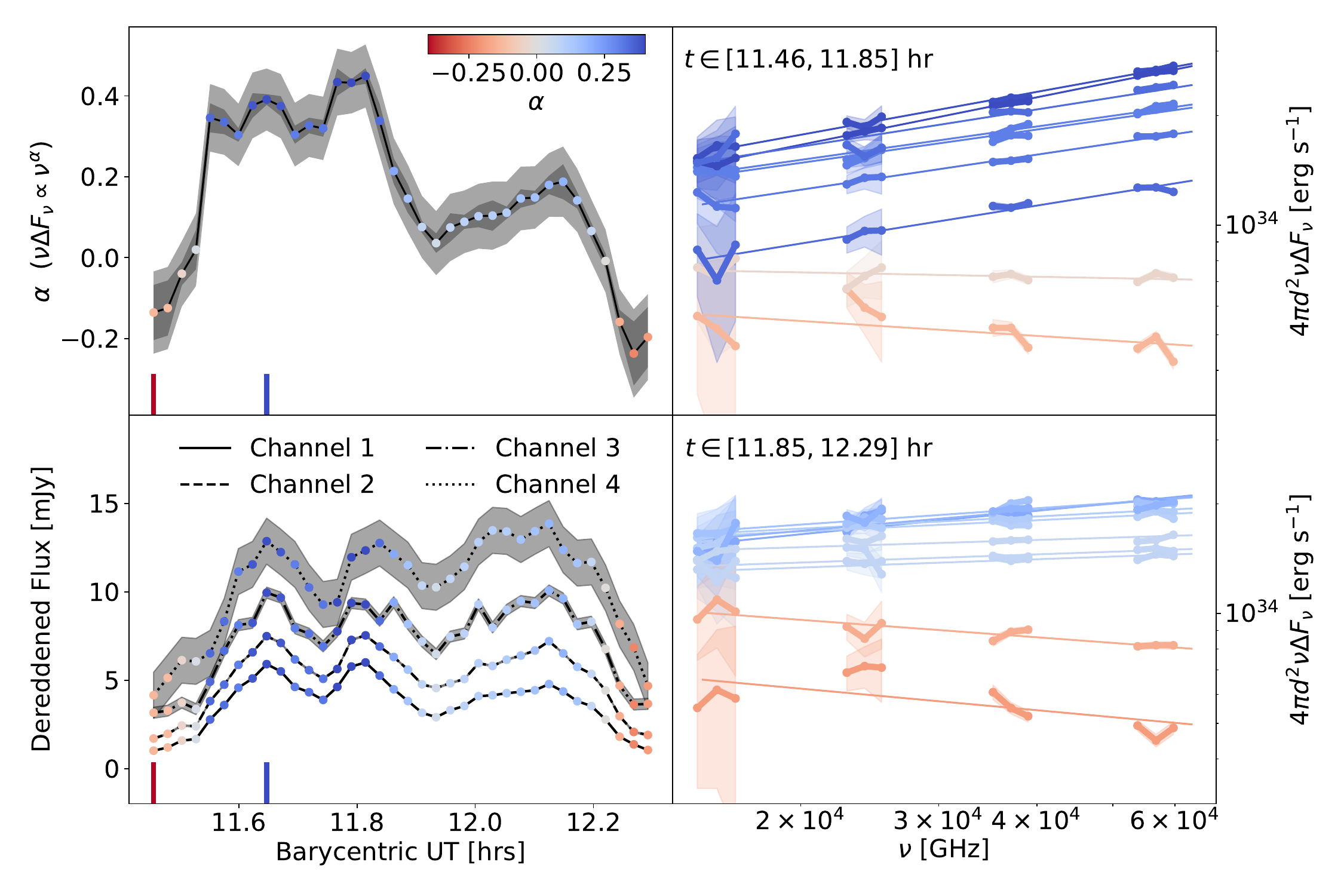}
            \caption{Temporal evolution of the mid-IR SED during the flaring event. \textit{Top Left}: Spectral index variations. The dark gray shaded region is the total statistical uncertainty, while lighter gray denotes the total statistical + systematic range. \textit{Bottom Left}: Intrinsic mid-IR light curves of \Sg\  in all four channels; the gray shaded region shows the statistical photometric noise in the measurement. The red and blue vertical bars denote times at which the time-resolved SEDs are plotted in Figure \ref{fig:mwl_sed}. \textit{Right panels}: Residual-extinction-corrected SEDs (see Appendix \ref{appx:residcorr}) at different times during the observation. Each MIRI/MRS channel has been broken up into three equally-wide frequency bins (connected points), and a best-fit power-law (solid lines) has been plotted. Shaded regions correspond to the statistical uncertainty in the measurement. The colors of data points in all panels correspond to the measured spectral indices.}
            \label{fig:timed_sed}
        \end{figure*}

        \Sg\  is known to be variable at all times in the near-IR, exhibiting red-noise properties with no (static) periodicity \citep{Do2009}. This red-noise-like behavior is most prevalent on $\lesssim$3-hour timescales and transitions to white noise on longer timescales \citep{Witzel2012, Witzel2018}. However, a long-standing observational question is how to demarcate a ``flare'' from the otherwise-continual IR variability. Flares have been phenomenologically defined as large-amplitude excursions in the IR and X-ray, on top of a continuously variable (IR-) quiescent state, which has been proposed by \cite{Do2009} and \cite{Dodds-Eden2011}. Specifically, \citet{Dodds-Eden2011} and \citet{GravityCollaboration2020flux} proposed a definition of a flaring state based on the near-IR flux distribution which is log-right-skewed, i.e., not log-normal. This non-log-normality was interpreted as evidence for two emission mechanisms at work, split into a log-normal quiescent emission process and power-law tail caused by flux contributed from the flaring process. This idea was challenged by other flux-distribution studies (\citealt{Witzel2012, Witzel2018}, and, to some extent, \citealt{GravityCollaboration2020flux}), which, while supporting non-log-normality, found that other distribution functions (such as a truncated power law, a double log-normal, or inverse-gamma distribution) could describe the flux distribution equally well. 
               
        Similarly, the near-IR spectral index has been used to argue for a distinct flaring state, where some authors have found a transition from steep spectral indices (as extreme as $\nu L_\nu \propto \nu^{-3}$ for quiescent fluxes, \citealt{Gillessen2006}) to shallower ones during bright states ($\nu L_\nu \propto \nu^{+0.5}$, \citealt{Hornstein2007}). While the spectral index during bright states is well-accepted \citep[e.g.,][]{Dodds-Eden2010, Ponti2017, Witzel2018, GravityCollaboration2021_xrayflare,Paugnat2024}, the low-flux-state spectral index is debated. This stems, in large part, from the limited atmospheric windows available to ground-based observations and their poor photometric stability \citep[see, e.g.,][for two approaches using IFU and imaging data]{Gillessen2006,Hornstein2007}. For instance, in an extensive statistical analysis of Keck light curves, \cite{Paugnat2024} estimated $1\sigma$ uncertainties $\Delta \alpha_{\rm{NIR}}
        \sim 1$--3 for low flux states and $\Delta \alpha_{\rm{NIR}} \sim 5$ at $3\sigma$-level.
    
        The existence of a distinct quiescence state is not trivial but relates to two competing ideas regarding how the near-IR and X-ray bright emission is generated, specifically whether flares are event-like \citep[e.g.,][]{Broderick2006_hotspot,Eckart2006,Gillessen2006,Dodds-Eden2010, Dodds-Eden2011, Ponti2017,GRAVITYCollaboration2018_orbital,GravityCollaboration2020_orbital,GravityCollaboration2021_xrayflare,Michail2021,Michail2023,Michail2024} or are statistical bright fluxes simply generated by a single stochastic red-noise process \citep[e.g.,][]{Do2009,Witzel2012,Meyer2014,Witzel2018,Witzel2021}. At the same time, \cite{vonFellenberg2023_sgra,vonFellenberg2024} showed that the question of event-like flares is disjunct from whether two emission processes are required \citep[for a general discussion of characterization of events in time series, see also][]{Scargle2020}. Specifically, they found that both event-like one- and two-state models match the near-IR flux distribution and the second and third order structure functions, which measure the variance and skewness in the light curves at any time lag. While simplicity arguments favor one-state models, near-IR time series measurements at a single frequency may be an insufficient tool to statistically rule out the possibility of two states in \Sg. 

        This Section extends the discussion by \cite{Gillessen2006}, which assessed whether a distinct flaring state exists based on a change in the spectral index at the flare's onset, i.e., a transition from a ``quiescent'' to flaring state. This is enabled by the high photometric stability of the JWST/MIRI light curve. Specifically, the observed spectral index changes abruptly by $\Delta \alpha_{\rm{MIR}}=0.33~\pm0.06_{\rm{stat}}~\pm0.11_{\rm{sys}}$\footnote{The photometric and extinction contribution of the systematic error reported here affect the absolute calibration but are negligible in the time series.} at 11:33:05 UT, corresponding to the perceived onset of the flare in the light curve (Figure~\ref{fig:timed_sed}). This observed behavior indicates a new, quantitative way to define a flare state: a quasi-instantaneous\footnote{Measuring the actual duration of the spectral change is limited by the 86-second cadence of the mid-IR light curves, meaning only an upper limit on its value is placed.} change in the spectral index.
        
        As the flare flux rises, the spectral index increases only modestly by $\Delta \alpha_{\rm{MIR}}=0.05\pm 0.04_{\rm{stat}}\pm0.11_{\rm{sys}}$, which implies no direct correlation of the spectral index with the increased flux. After the spectral index reaches its peak value of $\alpha_{\rm{MIR}}=0.45\pm 0.01_{\rm{stat}}\pm0.08_{\rm{sys}}$ at 11:48:51 UT, it drops by $\Delta \alpha_{\rm{MIR}} = -0.41\pm 0.03_{\rm{stat}}\pm0.11_{\rm{sys}}$ over ${\sim} 7~\mathrm{minutes}$. After this drop, $\alpha_{\rm{MIR}}$ changes by only $\Delta \alpha_{\rm{MIR}}= 0.15\pm 0.05_{\rm{stat}}\pm0.11_{\rm{sys}}$ over ${\sim} 10~\mathrm{minutes}$. As the flux drops further, the spectral index reverts to  values similar to those before the onset of the flare. 

        This abrupt onset can be linked to event-like particle acceleration in the accretion flow. For example, assuming that the observed luminosity transitions from a pre-flare state to an electron spectrum with a non-thermal tail, the flare onset is defined when the observed emission transitions into the accelerated spectrum. 
        This terminology may be misleading, as the underlying electron distribution before the flare may be non-thermal, for instance if the synchrotron-cooling break resides at lower frequencies before the onset of the flare, which may be more similar to the ``one-state'' concept. 
        At the same time, we do not rule out that the dominant electron distribution before and after the flare may be different altogether, more akin to the ``two-state'' concept. 
        
        In either case, adopting such a flare picture encodes a wealth of information about the accelerating mechanism at the transition time. 
        First, during the onset of the flare, the electrons have not yet substantially cooled, so the spectrum is dominated by that of the accelerated particles.
        If we assume the non-thermal electron distribution $n(E)$ powering the flare's synchrotron emission is power-law-distributed, i.e., $n(E)\propto E^{p}$, then the observed mid-IR spectral index, $\alpha_{\rm{MIR}}=0.39\pm0.01_{\rm{stat}}\pm0.08_{\rm{sys}}$, can be directly be related to $p$: $p_{\rm{electron}} = 2\alpha_{\rm{MIR}}-3= -2.2\pm 0.1_{\rm{stat}}\pm0.2_{\rm{sys}}$.

        Second, the timescale by which the spectrum changes to the accelerated spectrum provides insight into the dimensionality of the system. For instance, if magnetic reconnection is considered as the acceleration mechanism, two timescales are relevant: one to accelerate an individual electron and another to amass a sufficient current sheet to produce a significant population of non-thermal particles. The first timescale is on the order of a few $10\times r_{\rm{Larmor, B=40G}}/c \sim 5~\mathrm{m}/c$ and thus negligible with respect to the resolvable timescales with these observations. The second timescale is more uncertain as there is no consensus on the mechanism driving the acceleration. Still, low-density cavities in the accretion flow, so-called flux tubes, have received significant attention \citep[e.g.,][]{Ripperda2020, Ripperda2022, Dexter2020_flare,Dexter2020_model, Porth2021}. Specifically, \citet{Ripperda2022} and \citet{Zhdankin2023} have proposed that Rayleigh--Taylor instabilities (RTI) power continuous particle acceleration through magnetic reconnection in regions where low- and high-density plasma and the respective magnetic fields mix. \cite{Ripperda2022} found typical length scales on the order of $1~R_g$ for these regions (referred to as RTI fingers), which suggests a timescale for the formation of the structure accelerating the particles of $R_g/c\sim 20$ s.

        Our light curve is too coarsely sampled temporally to resolve such buildups. However, if the 86-s cadence is interpreted as an upper limit on the acceleration region size, it places a size constraint of $R_g/c \times 86~{\rm{s}} \lesssim4~R_g$, which holds for any acceleration mechanism assuming the region is causally associated. 

        The spectral index decreases after $\sim$20 minutes, consistent with a cooling synchrotron population in line with the discussion of \citetalias{vonFellenberg2025}. Still, the slight increase of the spectral index during the last peak (upper left panel of Figure \ref{fig:timed_sed}) indicates some electron re-injection or variability of the underlying electron distribution, indicative of a more complex injection profile than the simple single- and double-Gaussian profiles chosen in \citetalias{vonFellenberg2025}'s initial analysis. We defer explicit modeling of the electron injection profile to upcoming work.

        Third, the lack of spectral index correlation with the rising flank of the flare flux ($t\lesssim$11:39 UT) indicates that achromatic Doppler boosting may explain some of the variability, as has been argued in several other studies \citep[e.g.,][]{Hamaus2009,GRAVITYCollaboration2018_orbital, Haggard2019,GravityCollaboration2021_xrayflare, vonFellenberg2023_sgra}. 
        
        The above discussion focused solely on the spectral evolution during the flare. Due to the faintness of \Sg\ before the flare, only a short portion of the light curve when \Sg\  was still relatively bright allows reliable spectral index measurements. Extrapolating the mid-IR flux and spectral slope to the $K$-band gives an estimated dereddened flux density $F_{2.2~\rm{\mu m}}=0.71^{+0.25}_{-0.15}~\mathrm{mJy}$, corresponding to approximately the 30--40\% flux percentile found by \cite{GravityCollaboration2020flux}. Therefore, we cannot answer whether there is a strict flux-dependence of the spectrum nor provide estimates of the minimal spectral index at low-flux/quiescence states. Still, our measurements indicate the spectral index is redder for lower flux densities. This behavior is predicted by \citet[][their Fig.~19]{Witzel2018} using the statistical behavior of the near-IR dim-state flux distribution. Despite the limitations, our measurements show the spectral evolution is complex, and flux density alone may be an insufficient descriptor of the spectral state. 
        
        Finally, our measurements are fully consistent with values reported in previous literature, which have large uncertainties. For instance, the maximal range in $\alpha_{\rm{MIR}}$ values found in this study of $\Delta \alpha_{\rm{MIR}}\approx 0.6$ is within $1\sigma$ range of \cite{Paugnat2024}.

\section{Millimeter Polarization}\label{sec:pol_lcs}
    \subsection{Polarized Variability and Stokes $Q$--$U$ Loops}
        \begin{figure*}[h]
            \centering
            \includegraphics[width=1\linewidth]{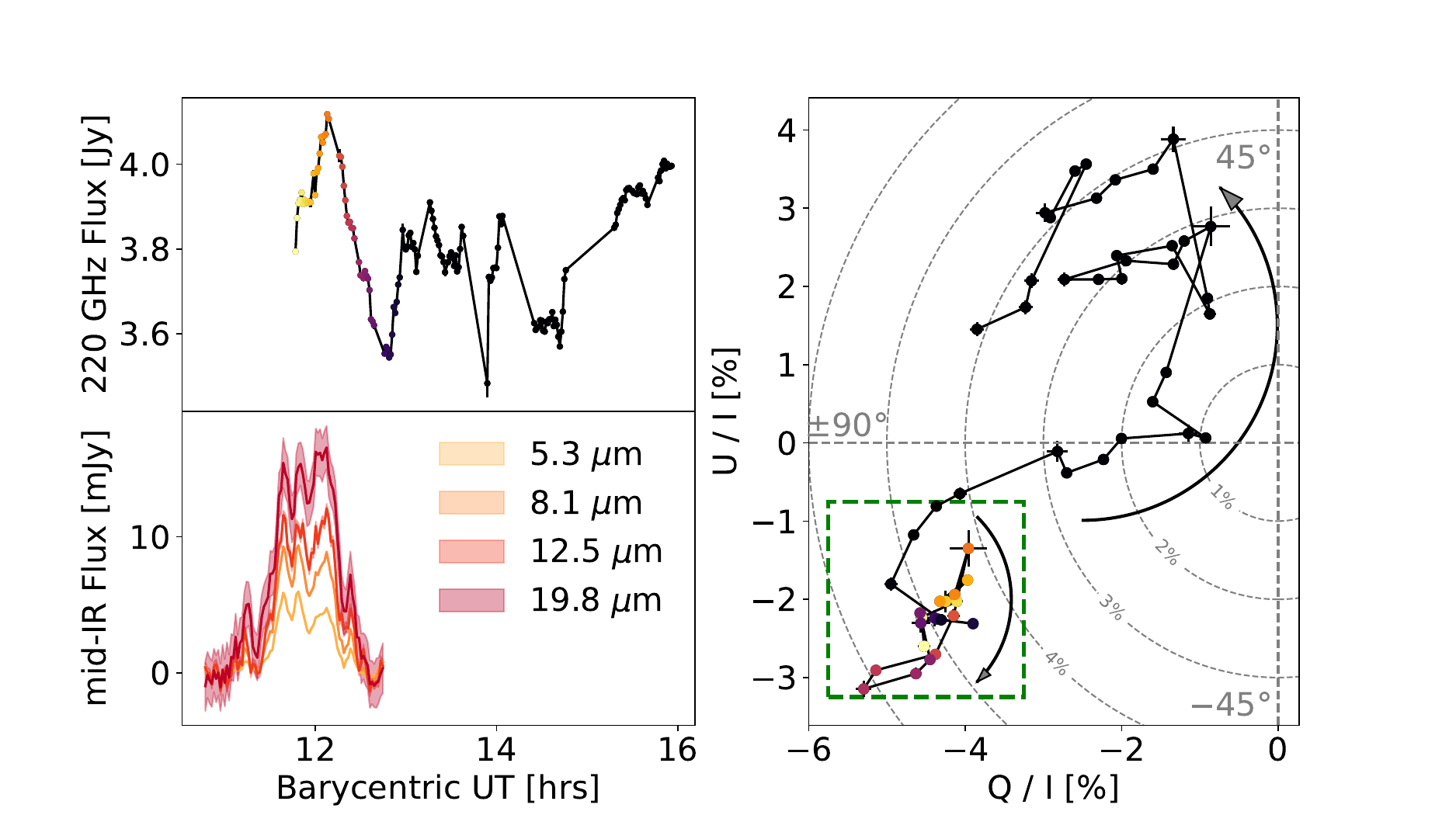}\\
            \includegraphics[trim = 0 0 0 1.cm, clip,width=1\linewidth]{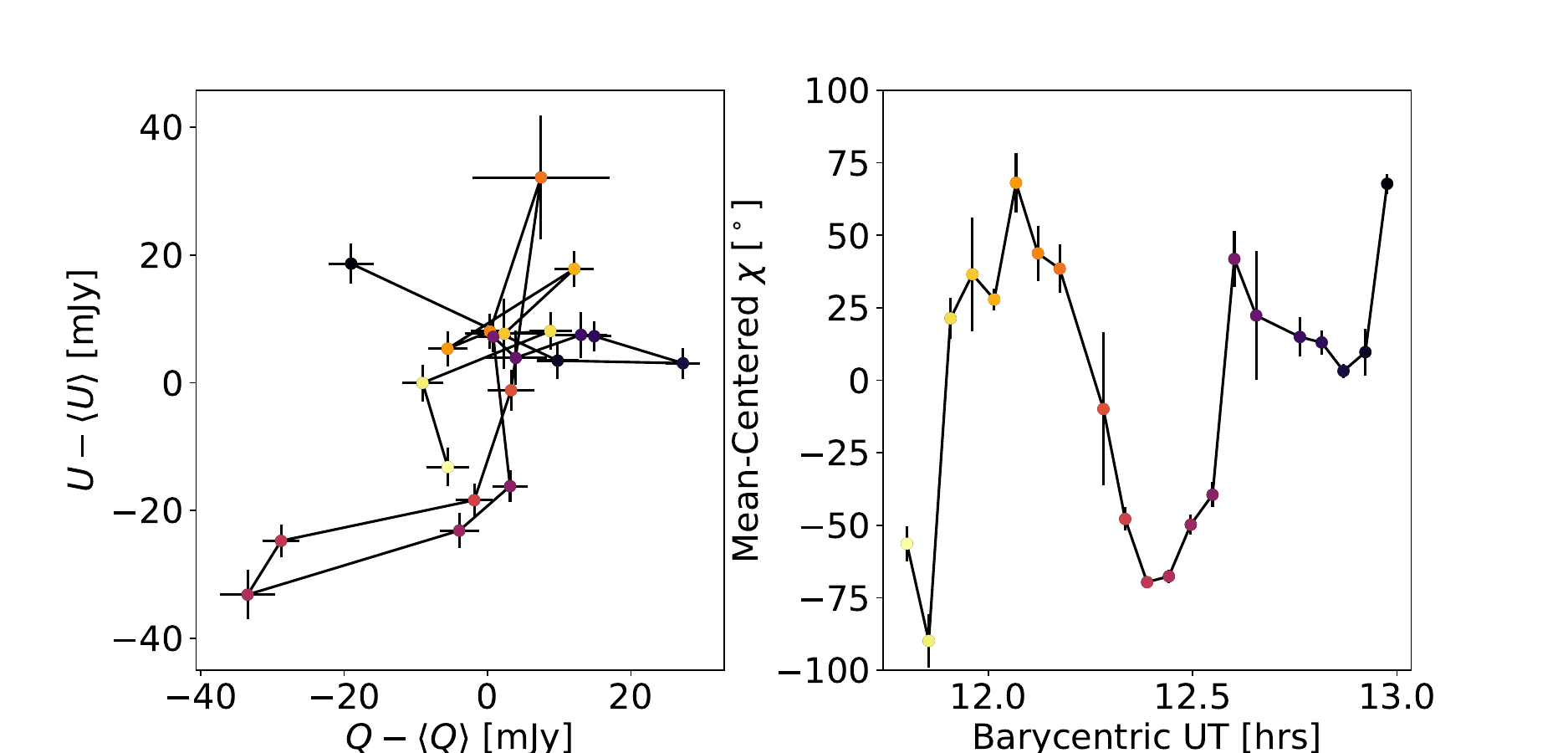}
            \caption{\textit{Top left}: Multiwavelength plot of SMA and dereddened JWST/MIRI light curves. Colored  points in the SMA light curve correspond to the barycentric time of each 60-second point. \textit{Top right}: Time-resolved mm Stokes $Q$--$U$ plot for \Sg\ obtained with the SMA at 192-second binning. Point colors correspond to the barycentric UT time in the top left panel. The green box corresponds to the zoom-in of the small $Q$--$U$ loop during the third mid-IR peak, and arrows show the overall orientation of the two observed loops. \textit{Bottom left}: Mean-centered zoom-in of the clockwise-oriented $Q$--$U$ loop before 13:00 UT at 192-second binning during the JWST/MIRI flare. \textit{Bottom right}: EVPA of the mean-centered Stokes $Q$--$U$ loop as a function of time, colored with the same scheme as the lower right panel.}
            \label{fig:qu}
        \end{figure*}

        The top right panel of Figure \ref{fig:qu} shows the temporal evolution of the Stokes $Q$ and $U$ light curves during the SMA observation. At the outset of the mm peak ($t\approx$12:00 UT), corresponding to the third mid-IR peak (top left panel), the linear polarization averages $\sim$5\% with an EVPA ${\approx}-74^\circ$. The lower left panel shows an inset of the $Q$--$U$ plot during the final mid-IR peak but with the mean Stokes $Q$ and $U$ values of this time range subtracted to remove the thermal contribution from the accretion flow. This mimics near-IR polarization plots \citep[e.g.,][]{GRAVITY_collab2023_polariflares} that are not blended with a polarized thermal component, thus directly probing the hotspot-induced $Q$--$U$ variability. While the ``loop'' is evidently distorted from circular shape, overall, there is a small, clockwise-(CW)-oriented polarization with a period of $\sim50$ minutes. The sense of rotation is more clearly present in the bottom right panel, which shows the EVPA of the mean-centered Stokes $Q$ and $U$ parameters. The overall negative change in EVPA, corresponding to CW rotation, hints at orbital-induced variations in the polarimetry. These properties match the senses of rotation and timescales of previous mm and near-IR polarization studies. 
        
        \citetalias{vonFellenberg2025} could not directly associate the $\Delta F_{220~\rm{GHz}}=300$ mJy mm variability with the mid-IR flare, but the polarimetric data now argue for a physical relationship between the mid-IR and mm peaks. First, assuming this loop is produced by an orbiting hotspot, the period of the polarization loop gives a direct estimate of the orbital radius.
        If the hotspot's magnetic field is poloidally dominated \citep[as observed in GRMHD simulations of magnetic-flux eruptions;][]{Ripperda2022} and the orbit is approximately Keplerian, a polarization period $P_{\rm{{pol}}}\approx50$ minutes yields an orbital radius of ${\sim}8~R_g$ \citep[][their Appendix B]{GRAVITYCollaboration2018_orbital}. This is consistent with the time between the first and the third peak, which was interpreted as Doppler boosting caused by orbital motion on a similar radius in \citetalias{vonFellenberg2025}. Two caveats are that the flux tube’s orbital velocity may be sub-Keplerian \citep[as suggested in GRMHD models by][]{Porth2021} or that the accretion flow may be more complex \citep[e.g., in wind-fed accretion models;][]{Ressler2020}, where the disk and flaring region magnetic fields are not orthogonal. In the sub-Keplerian case, the derived orbital radius is a lower limit that is otherwise poorly constrained. A forthcoming study will better constrain its orbital velocity but is outside the scope of the current work.

        Secondly, we expect if the polarization signature is caused by an orbiting hotspot, and the emission is the same as the IR emission, its intrinsic linear polarization ($p_{\rm{int}}$) should be comparable to typical polarization fractions found in IR flares \citep[few $\times10\%$ to $\sim$30\%;][]{GravityCollaboration2020_polariflares}.
        We estimate $\Delta I_{\rm{flare}} \approx 300 ~\mathrm{mJy}$ and $\Delta p_l \approx 50~\mathrm{mJy}$ (Figure \ref{fig:qu}), yielding $p_{\rm{int}} \sim 20\%$, consistent with the above-mentioned values. The flare's high intrinsic linear polarization is also directly seen in the polarization percent's change between the final mid-IR peak ($\sim$5\%) and mm ``quiescence'' ($\sim$2\%; third panel of Figure \ref{fig:lightcurves}). Similar polarization fractions have been detected in other mm polarization loops  \citep[$\lesssim50\%$;][]{Wielgus2022,Michail2023}.
        
        A strong, testable prediction of this mm-to-mid-IR connection in future analyses is a relationship between the ``size'' of the $Q$--$U$ loops and the mid-IR spectrum. Brighter and/or softer mid-IR spectra will produce brighter mm variable emission ($I_{\rm{flare}}$) from the same non-thermal population of electrons, while dimmer and/or harder mid-IR spectra will produce less mm emission.\footnote{This simple framework cannot as easily describe the intermediate cases where the mid-IR emission is dimmer and softer or brighter and harder, which will require an analytic model.} Assuming the other hotspot properties such as their intrinsic linear polarization fraction $p_{\rm{int}}$, remain roughly constant, the size of the orbit-induced $Q$--$U$ loop is directly proportional to the mm polarized flux (i.e., $p_{\rm{int}}I_{\rm{flare}} \sim \sqrt{Q^2_{\rm{flare}} + U^2_{\rm{flare}}}$). Thus, the relationship between the size of the mm $Q$--$U$ loop and mid-IR flare are directly verifiable.
        
        
        After the mid-IR flare subsided, a counterclockwise (CCW) loop began near $t\sim13$:00 UT. Two other instances of CCW loops observed in \Sg's light curves are published: one at 230 GHz \citep{Marrone2006} and the other at 86 GHz  \citep{Wielgus2024}. In all three cases, the CCW loop shows a period of $\sim$3 hours. \cite{Wielgus2024} noted that if orbital motion is invoked, this period corresponds to an orbital radius of ${\approx}20~R_g$, albeit in the opposite orientation of other polarimetry results. From a theoretical perspective, \citet{Grigorian2024}, \citet{NajafiZiyazi2024} and \citet{Ricarte2025} analyzed mm $Q$--$U$ loops in general relativistic magnetohydrodynamic (GRMHD) simulations of \Sg\ and found that CCW-oriented loops are occasionally produced. \citet{Grigorian2024} and \citet{NajafiZiyazi2024} found CCW loops appearing during ``flaring'' episodes and following the orbital orientation of the accretion flow. \citet{Ricarte2025}, on the other hand, found occasional periods of CCW loops outside of flaring episodes and opposite to the accretion flow orbit albeit on shorter timescales ($\sim$few$\times10$ minutes). 
        Observationally, however, the common 3-hour period between all three known CCW loops tends to suggest a common formation mechanism.

    \subsection{Rotation-Measure Variability}\label{ssec:rm}
        Given the substantial change in the linear polarization, we searched for similar changes in the rotation measure (RM), which probes the integrated electron-density weighted line-of-sight magnetic field (i.e., RM $\propto\int n_{\rm e}\vec{B}\cdot d\vec{l}$). The relationship between RM and the EVPA is:
        \begin{equation}
            \chi(\lambda) = \chi_0 + \rm{RM}~\lambda^2,
            \label{eq:rm}
        \end{equation}
        where $\chi(\lambda)$ and $\chi_0$ are the EVPA at any $\lambda$ and the intrinsic EVPA, respectively. Fitting the LSB and USB data at each timestep cannot determine $\chi_0$ because of the small bandwidth but does constrain RM\null.  Before 13:15 UT, \Sg's (inverse-variance-weighted) average $\mathrm{RM}=(-2.57 \pm0.09)\times10^5$ rad m$^{-2}$. Averaging all the time steps after 13:15 UT gives $(-3.85\pm0.12)\times10^{5}$ rad m$^{-2}$, suggesting a statistically-significant change. The same analysis for J1733 gives $(-4.85 \pm 2.48)\times10^5$ rad m$^{-2}$ and $(0.66 \pm 1.51)\times10^{5}$ rad m$^{-2}$, respectively, $\la2\sigma$, whereas \Sg's change is $\approx9\sigma$. Including residual thermal effects (Appendix \ref{appx:thermalrm}) yields a still-significant ${\sim}4.5\sigma$ RM change for \Sg. Therefore, \Sg\ experiences quick ($\sim$30-minute; see the right panel of Figure \ref{fig:lightcurves}), correlated changes in both the linear polarization and RM after the final mid-IR peak. This timescale is much shorter than the mm synchrotron cooling time \citep[$\sim3$ hours at 220 GHz in a 30 Gauss field;][]{Dodds-Eden2010}, and inefficient particle escape from the turbulent zone powering the flare \citep[][]{Kempksi2023,Lemoine2023} should largely contain electrons within the boundaries, rendering escape timescales much larger than the observed polarization and RM changes. Hence, the shift in the value of the RM tends to suggest the end of the mid-IR flare. 

\section{Summary and Conclusions}
    This paper presents the first time-resolved mid-IR SED of \Sg\ during a flare. The observed light curve was first published by \cite{vonFellenberg2025}, \citetalias{vonFellenberg2025}, but dereddening with the new extinction measurements from \citepalias{vonFellenberg2025_extinction} and our new method to calculate aperture corrections for non-dithered MIRI/MRS observations allow comparison with physical models.
    \begin{enumerate}
        \item The maximal mid-IR spectral index $\nu \Delta F_\nu\propto\nu^{+0.45\pm0.01_{\rm{stat}}\pm0.08_{\rm{sys}}}$ matches the near-IR spectral index \citep[e.g.,][]{Hornstein2007, Ponti2017, Paugnat2024} $\nu F_\nu\propto\nu^{+0.5}$ in bright states.
        \item We identify a sudden increase in the mid-IR spectral index ($\Delta\alpha_{\rm{MIR}}=0.33\pm0.06_{\rm{stat}}\pm0.11_{\rm{sys}}$) at the onset of the flare, which we interpret as the transition to an accelerated (hard), non-thermal electron distribution. If confirmed in future observations of mid-IR flares, this offers a novel, spectral-index-based definition of flare onset. 
        \item If the sudden spectral-index increase represents flare onset, the 86-second upper limit on its duration yields a size of ${<}86~{\rm{s}}/c \lesssim 4~R_g$ for the accelerating region.
        \item Before the near-instantaneous spectral index change, spectral indices were $\nu \Delta F_\nu\propto\nu^{-0.2\dots0.0}$. However, the time range of significant low-flux-density spectral measurements is small in the analyzed dataset, prohibiting a general statement on the mid-IR quiescent spectral range.
        \item The absolute spectral index analysis presented here aligns fully with the relative evolution reported in \citetalias{vonFellenberg2025}, reinforcing the validity of the synchrotron-cooling results established in that earlier work.
    \end{enumerate}
    
    In addition, new mm linear-polarization measurements observed with the SMA 
    extend the total-intensity measurements reported in \citetalias{vonFellenberg2025}. The total intensity measurement showed a $\Delta F_{220~\mathrm{GHz}} \approx 0.3~\mathrm{Jy}$ peak in the light curve on top of mean flux ${\sim}3.8~\mathrm{Jy}$. In \citetalias{vonFellenberg2025}, this peak could be linked to mid-IR variability if the number density of the mid-IR-emitting electrons was sufficiently large. However, absent extinction correction, \citetalias{vonFellenberg2025} could not constrain the electron density from the mid-IR data alone. Now, extrapolating the dereddened mid-IR flare spectrum with a power law shows that the connection between the mid-IR flare and mm peak is indeed plausible.
    
    Further, significant variability in the mm linear polarization fraction, angle, and rotation measure (RM) during the evolution of the mid-IR flare, show:
    \begin{enumerate}
        \item There was a (distorted) clockwise-oriented $Q$--$U$ loop with a $\sim$50-minute period that occurred during a stretch of higher linear polarization, both of which coincided with the third mid-IR flare peak.
        \item The total variability of polarized flux was ${\sim} 50~\mathrm{mJy}$, suggesting a polarization fraction of 10--20\% if we assume the $\Delta F_{220~\rm{GHz}}=300~\mathrm{mJy}$ Stokes $I$ variability was caused by a localized emission region linked to the mid-IR flare. This polarization fraction is consistent with the typical polarization fraction observed in the near-IR \citep[e.g.,][]{Shahzamanian2015,GravityCollaboration2020_polariflares} and with the magnitude of $Q$--$U$ loop measured by \cite{Wielgus2022}.
        \item We measured a ${\sim}4.5\sigma$ change in the RM after the mid-IR flare ended and the mm light curve had dropped to its minimum. 
    \end{enumerate}

    The last point may indicate that a change in the electron number density and/or the magnetic field configuration occurs at the end of the flare. If flares originate from low-density flux tubes where RTI drives magnetic reconnection \citep[e.g.,][]{Ripperda2022,Zhdankin2023}, this may indicate that the flux tube is disrupted (e.g., by mixing instabilities at the boundary between the tube and the accretion disk) at the end of the flare \citep[e.g.,][B. Seefeldt-Gail et al., in prep.]{Ripperda2022}. This may explain the absence of additional $Q$--$U$ loops and prolonged mm variability, which would be naively expected if only synchrotron cooling is considered. In the mm, synchrotron cooling timescales are on the order of several hours for the magnetic field strengths explored in \citetalias[][]{vonFellenberg2025}.
    However, we observes a 3-hour counterclockwise $Q$--$U$ loop after the mid-IR flare subsided. Such counterclockwise $Q$--$U$ loops were previously observed at 86 GHz \citep{Wielgus2024} and 230 GHz \citep{Marrone2006}. GRMHD simulations can produce hour-long counterclockwise $Q$--$U$ loops during flaring events when the accretion flow's orbital orientation is in the same direction \citep{Grigorian2024, NajafiZiyazi2024}, but their lifetimes are limited to a few tens of minutes outside of flaring events when in the opposite sense of direction as the accretion flow \citep[][]{Ricarte2025}.
    GRMHD simulations with purely thermal electrons show that both clockwise and counter-clockwise polarization rotations can arise spontaneously from internal Faraday rotation and turbulence, independent of flaring activity \citep{Ricarte2025}. The observed post-flare CCW loop in our data may therefore not trace bulk orbital motion but instead reflect polarization fluctuations in the cooling or reconnection-disrupted plasma. 
    Therefore, it may support the idea that only a transient fraction of the mm (polarization) variability is of non-thermal IR and X-ray origin. Still, it may also serve as a cautionary note that the mm variability is likely more complicated than alluded to here, as the non-thermally-produced variability must compete with the overall thermal mm variability having the same order of magnitude. Further analysis of simultaneous full-polarization mm and mid-IR/X-ray observations are required to discern the exact contribution of non-thermal contribution at mm wavelengths.

\appendix
\section{Millimeter Polarization Data and Comparisons}
    \subsection{Polarization and Spectral Index Conventions}\label{ssec:sma_data_conv}
        \begin{figure}
            \centering
            \includegraphics[width=0.45\linewidth]{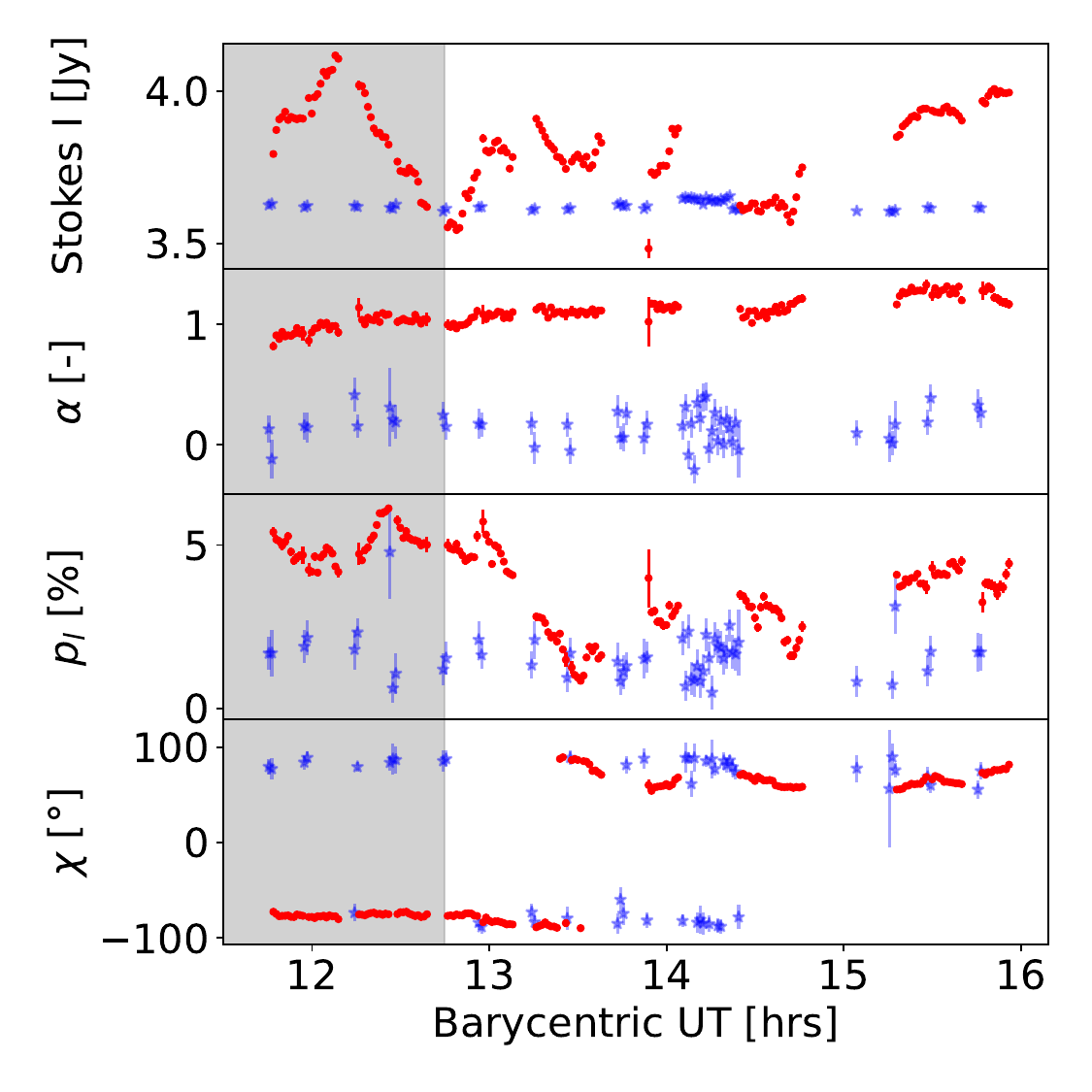}
            \includegraphics[width=0.45\linewidth]{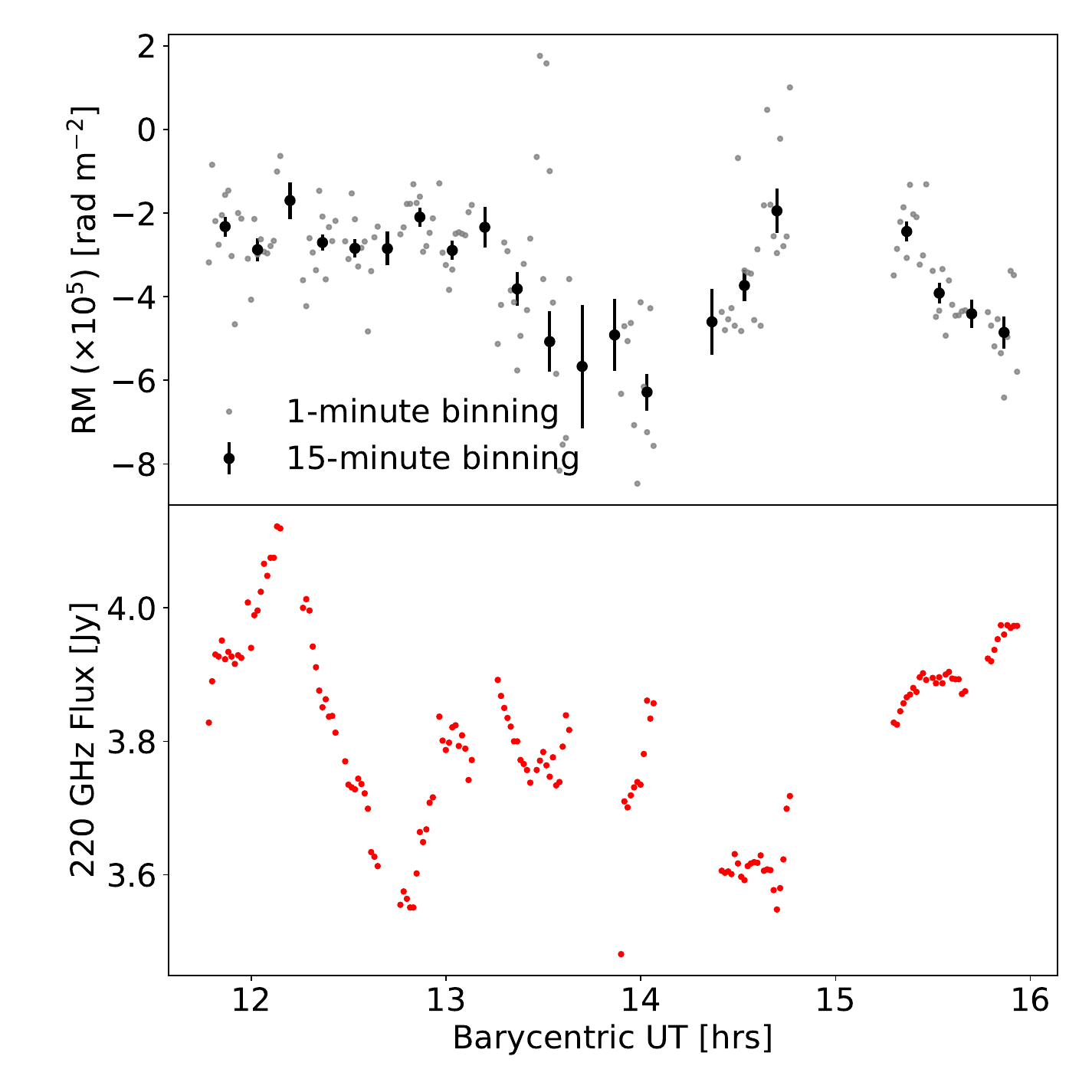}
            \caption{\textit{Left}: Stokes $I$, spectral index, debiased linear polarization percent, and EVPA light curves for \Sg\ (red) and the main gain calibrator J1733 (blue) on 2024 April 6. The flux density of J1733 is shifted up by 2.75 Jy. The gray shaded region denotes the time range where JWST/MIRI was observing \Sg. \textit{Right}: RM variations as a function of time. Native resolution (1-minute) values have been binned to 15 minutes using uncertainty-weighted averages. The SMA light curve is also plotted in the bottom panel.}
            \label{fig:lightcurves}
        \end{figure}
        \noindent The EVPA is defined as
            \begin{equation}
                \chi = \dfrac{1}{2}~\rm{arctan2}\left(\dfrac{{U}}{{Q}}\right),
            \end{equation}
        where $\rm{arctan2}$ places the angle in the correct quadrant. The de-biased linear polarization fraction is calculated with
            \begin{equation}
                p_l = \sqrt{p_{b,l}^2 - \sigma_{p_l}^2},    
            \end{equation}
        where $p_{b,l}$ is the (biased) polarized fraction:
            \begin{equation}
                p_{b,l} = \dfrac{\sqrt{Q^2 + U^2}}{I},
            \end{equation}
        and $\sigma_{p_l}$ is the uncertainty of $p_{b,l}$. (See Appendix of \citealt{Michail2023} for further details.) The mm spectral index is calculated between the LSB and USB light curves. The mm Stokes $I$ light curve and polarization products are plotted in Figure~\ref{fig:lightcurves}.
    
    \subsection{Comparing Observation-Averaged Polarization with AMAPOLA}\label{ssec:average_pol_results}
        \begin{figure*}[h]
            \centering
            \includegraphics[width=1\linewidth]{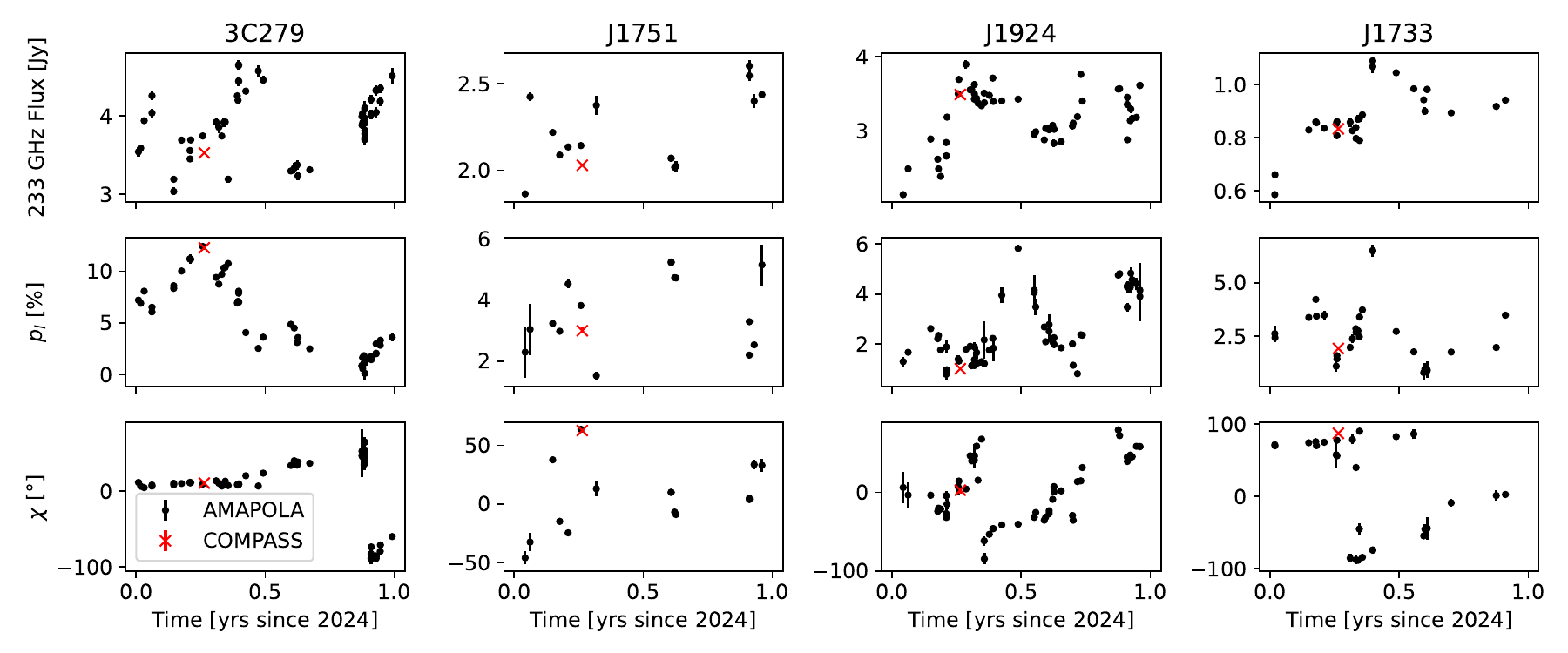}
            \caption{Comparisons of Stokes $I$, linear polarization, and EVPA (top, center, and bottom rows, respectively) for the four calibrators in this track using AMAPOLA (ALMA, black dots) and SMA (calibrated through COMPASS and CASA, red crosses) during 2024, showing that the calibrator properties are consistent with each other. The RMS-based error bars for SMA are plotted but smaller than the marker size.}
            \label{fig:compass_amapola}
        \end{figure*}
        
         Given the novelty of the CASA-based SMA polarization calibration technique used here, we compared the observation-averaged polarization properties of the four calibrator sources with AMAPOLA results at 233 GHz to ensure consistency. We imaged each calibrator using \texttt{TCLEAN} with the Multi-Term Multi-Frequency Synthesis (MTMFS) deconvolver (\texttt{deconvolver=`mtmfs'}) in all four Stokes parameters (\texttt{stokes=`IQUV'}) using two frequency terms (\texttt{nterms=2}) at a reference frequency of 233 GHz (\texttt{reffreq=`233GHz'}). Our chosen maps have pixel scale 0\farcs25 (\texttt{cell=`0.25arcsec'}) and size 256 pixels (\texttt{imsize=256}), which we interactively masked and cleaned to a noise level of $3\sigma$ (\texttt{nsigma=3.0}). No Stokes $V$ emission was detected in any of the calibrators, and all of the calibrators appear as point sources in the final images. Because none of the sources is resolved, we report the flux density in the Stokes $I$, $Q$, and $U$ planes as the central pixel's value. (No primary beam correction is needed because all the sources were at the phase center.) We estimated the statistical uncertainty as the RMS of the residual within a $5\times5$ pixel aperture ($1\farcs25\times1\farcs25$) centered on the central pixel. We calculated the debiased polarization percent, EVPA, and their uncertainties using the standard equations presented in Section~\ref{ssec:sma}. Figure \ref{fig:compass_amapola} shows the comparison with 2024 AMAPOLA results, which demonstrate that the CASA-based SMA polarization reduction path is consistent with ALMA results.

    \subsection{Estimating Rotation Measure Thermal Noise}\label{appx:thermalrm}
    
        \begin{figure}[h]
            \centering
            \includegraphics[width=0.5\linewidth]{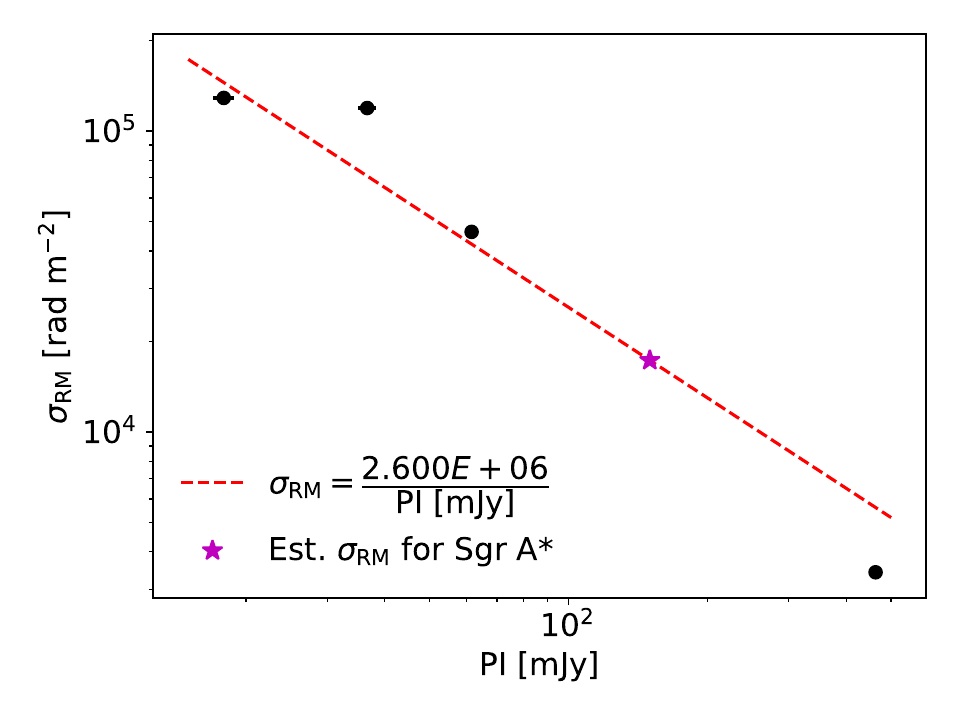}
            \caption{Estimate of the thermal RM noise present in the SMA data ($\sigma_{\rm{RM}}$) as a function of polarized flux density (PI) for the four calibrators (black points). The best-fit $\sigma_{\rm{RM}}\propto\rm{PI}^{-1}$ is plotted in red, and the purple star is the estimated thermal RM noise for \Sg.}
            \label{fig:systematicrm}
        \end{figure}
        
        We assessed the thermal noise on \Sg's derived RM using the other calibrators in the SMA observation. However, because 3C~279, J1751, J1733, and J1924 have different polarized fluxes, we cannot make a direct comparison to \Sg. Therefore, we derived a scaling relation between the thermal RM noise $\sigma_{\rm{RM}}$ and the average polarized flux ($\rm{PI}$) for the calibrators. Equation \ref{eq:rm} defines the relationship between RM and the measured EVPAs; therefore, $\sigma_{\rm{RM}}\propto\sigma_{\chi}$. Further, \citet{Montier2015} showed that $\sigma_{\chi}$ is related to the noise in the measured polarization fraction ($p$) and its uncertainty ($\sigma_{p}$): $\sigma_{\chi}\propto\sigma_p/p$. Finally, from the definition of PI, ${\rm{PI}}=pI$, where $I$ is the Stokes $I$ value. Disregarding $\sigma_I$ (because $\sigma_I\ll\sigma_p$), $\sigma_{\rm{PI}}\propto \sigma_pI$. Combining these scalings,
            \begin{equation}
              \sigma_{\chi}\propto\sigma_{\rm{RM}}\propto\dfrac{\sigma_p}{p} = \dfrac{\sigma_{PI}/I}{p} = \dfrac{\sigma_{\rm{PI}}}{\rm{PI}}\implies\sigma_{\rm{RM}}\propto\left(\rm{PI}\right)^{-1}
            \end{equation}
        We used this relationship to estimate $\sigma_{\rm{RM}}$ for \Sg\ (Figure \ref{fig:systematicrm}). \Sg\ had an average polarized flux of $\sim$150 mJy during this observation corresponding to $\sigma_{\rm{RM}}=0.17\times10^{5}$ rad m$^{-2}$. This residual thermal noise is subdominant to the observed shift in RM.
        
\section{MIRI/MRS Aperture Corrections for Non-Dithered Observations}\label{appx:apcorr}
    \begin{figure}[h]
        \centering
        \includegraphics[width=0.75\linewidth]{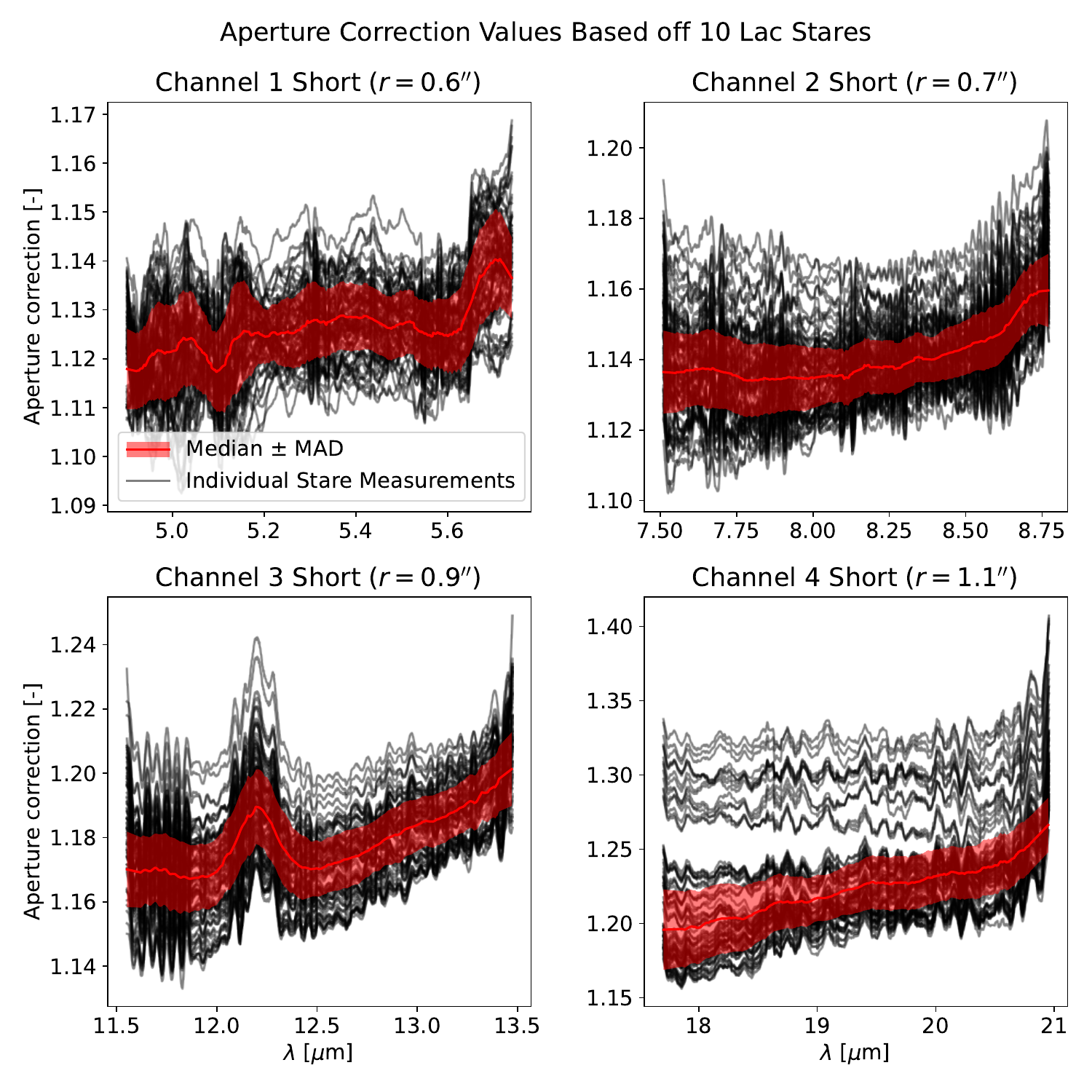}
        \caption{Derived aperture corrections for non-dithered observations of the star 10 Lac in circular apertures with channel-dependent radii. Gray shows the individual solutions for each of the 72 pointings. The red solid line is the smoothed median solution for the short gratings in each of the four channels, and the shaded region is the median absolute deviation (MAD) of the solutions.}
        \label{fig:apcorr_solutions}
    \end{figure}
    
    Time-resolved MIRI/MRS measurements require non-dithered observations: a single pointing and acquisition sequence per visit. While this mode produces high-time-resolution light curves with exceptional sensitivity in the mid-IR, obtaining calibrated flux densities presents a unique photometric challenge because dithering is required to fully sample MIRI/MRS's point-spread function (PSF)\null. MIRI/MRS is well-known to have a complicated PSF that produces both spatial and spectral features (resampling noise, \citealt{Law2023}, and fringing, \citealt{Argyriou2020}, respectively), which dithering suppresses. The malformed PSF is most readily seen in undithered point-source observations, where the source may be reconstructed as not pointlike. This issue, coupled with the loss of $\sim$5--10\% of the source's flux outside of MRS's field of view \citep{Law2025}, motivated us to develop an all-encompassing calibration to marginalize over the source's detector position and account for aperture corrections. 

    We used Cycle 2 MIRI/MRS calibration data of the star 10 Lacertae from PID 3779 (PI: D. Gasman; \citealt{2023jwst.prop.3779G}), originally observed to characterize spatially varying calibration solutions necessary for point sources. The data consist of channel 1 through 4 observations in the \texttt{SHORT}, \texttt{MEDIUM}, and \texttt{LONG} gratings. Each channel and grating was observed with two four-point dithers using a $3\times3$ pixel mosaic; in total, each channel+grating combination has 72 individual pointings. This analysis used only the \texttt{SHORT} observations.

    We processed RATE files through the standard Level 2 JWST pipeline (version 1.17) with the two-dimensional residual fringing correction turned on. We took two different reduction paths in the Level 3 pipeline, however: first, we piped all 72 pointings in each channel's \texttt{SHORT} grating together to produce four highly-sampled spectral cubes. This cube fully samples the PSF to calibrate the stares against. Then, to mimic our MRS observations, we processed each pointing separately to obtain estimates of the undersampled-PSF on 10 Lac. 

    We determined the ``infinite-aperture'' per-spectral-plane flux of 10 Lac on the fully-dithered spectral cubes using the astropy \texttt{photutil} package. The package calculated the background-subtracted photometry in a circular aperture centroided on 10 Lac and using the \texttt{MeanBackground} function to estimate the background. The radii in each channel ranged from 0.5 to 8.5 pix in steps of 0.5 pix to calculate the photometric growth curve. We extrapolated the total flux of the source by fitting the asymptotic function:
        \begin{equation*}
            f(n) = A\left(1-e^{bx}\right),
        \end{equation*}
    where $A$ is the ``infinite-aperture'' flux and $b$ is a scaling constant with unit $[\rm{pix}]^{-1}$. To correct for the known spectral leakage in the channel 3 \texttt{SHORT} filter from the channel 1 \texttt{MEDIUM} grating \citep{Argyriou2023a}, we followed the logic in the \texttt{SpectralLeakStep}\footnote{\url{https://jwst-pipeline.readthedocs.io/en/latest/api/jwst.spectral_leak.spectral_leak_step.SpectralLeakStep.html}} pipeline function. 
    
    To investigate the range of aperture corrections due to the undersampled PSF, we adopted a per-channel aperture radius as our standard extraction region and did not do any background subtraction. (The MIRI/MRS aperture corrections used in the pipeline are calculated differently in that a wavelength-dependent radius is used; \citealt{Law2025}.)
    The aperture sizes were chosen based on differential images of \Sg\ to maximize the aperture size (reducing the effect of non-dithered PSF artifacts) while excluding other variable or bright sources (Section \ref{ssec:jwst}). 
    
    The per-pointing aperture correction for each spectral plane was calculated as the quotient between the infinite-aperture flux ($A$ above) and the measured flux in our chosen aperture. The full set of aperture corrections is shown in gray in Figure~\ref{fig:apcorr_solutions}, where we have median filtered and smoothed each pointing using a first-order Savitzky--Golay filter with a window size of 21 channels. We calculated the median and median absolute deviation (MAD) over all pointings, again smoothed the results with a first-order Savitzky--Golay filter with a window size of 51 channels, and adopted the results as the wavelength-dependent aperture corrections and their uncertainties. These adopted corrections are shown in red in Figure~\ref{fig:apcorr_solutions}.

    Channel 3 has an excess aperture correction of $\approx$1--2\% caused by residual spectral leakage. This effect is generally within the MAD and subdominant to other sources of error in this broader analysis, and we have not tried to ameliorate it further. The larger-correction branch in the channel 4 results, differing by about $10\%$, correspond to files where 10 Lac is oblong in the reconstructed data. However, these do not correlate with position on the sky or detector or a single bad set of pointings. More accurate estimates of non-dithered aperture corrections, especially in channel 4, will require a set of observations where the calibration source is near the center of the rebuilt spectral cube.

    \section{Residual Extinction Correction and Uncorrected SEDs}\label{appx:residcorr}
    
        \begin{figure*}[h]
        \centering
            \includegraphics[width=0.75\linewidth]{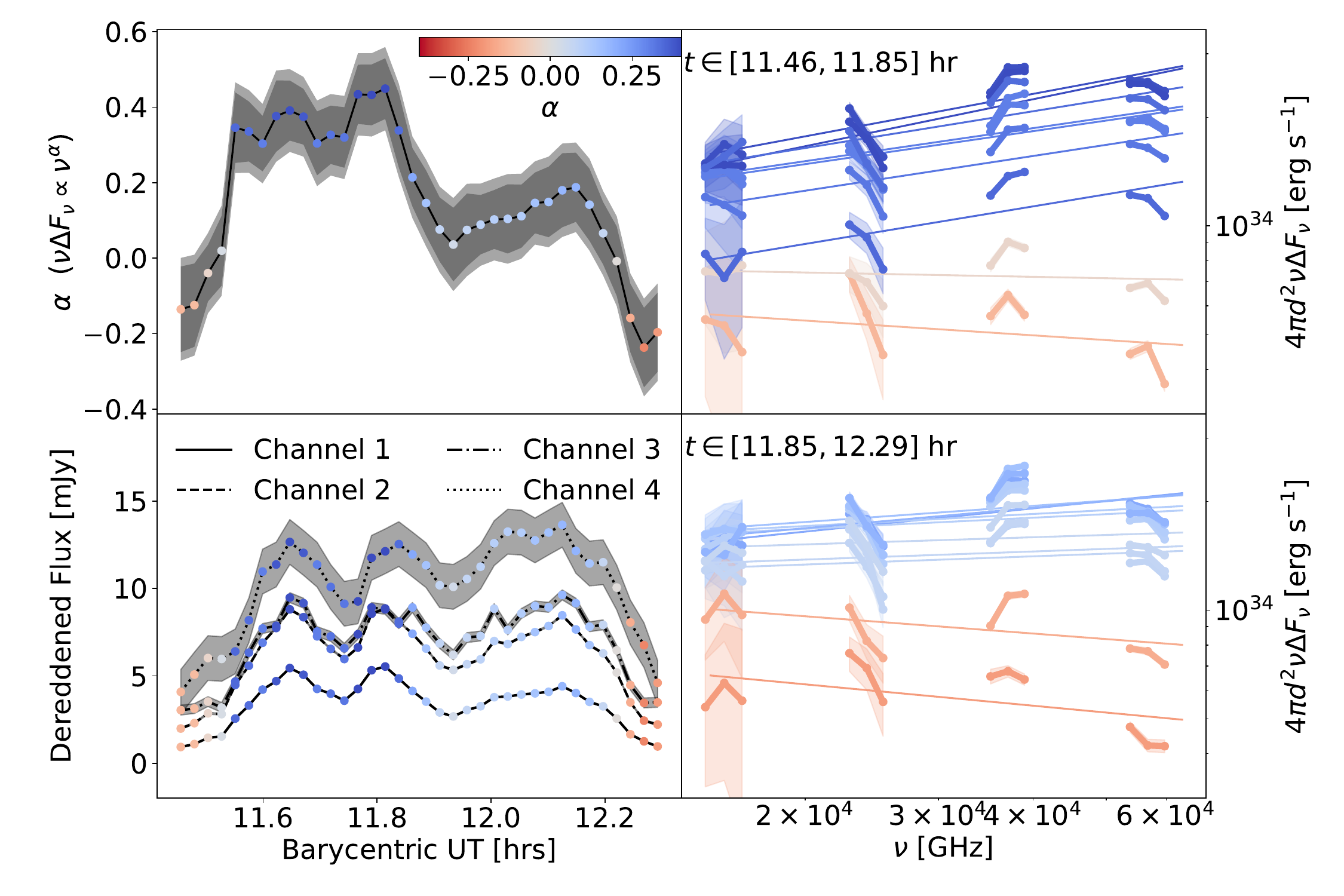}
            \includegraphics[width=0.55\linewidth]{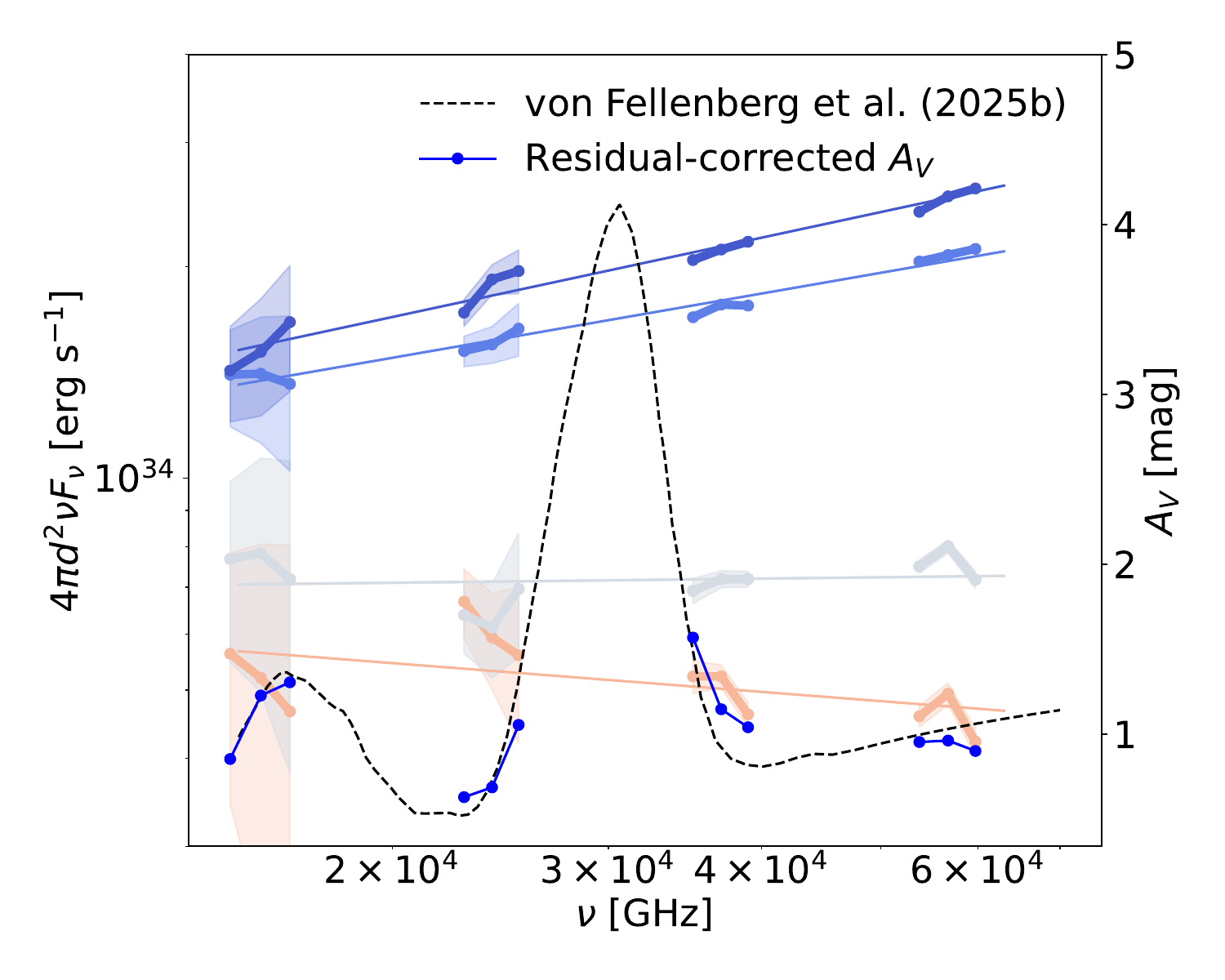}
            \caption{\textit{Top}: Same as Figure \ref{fig:timed_sed} but without the residual extinction correction applied. \textit{Bottom}: The extinction curve published in the \citetalias{vonFellenberg2025_extinction} (black) and the residual-corrected extinction curve (blue) are compared. The residual extinction was calculated assuming \Sg's SED is a power law during the flare. A few measurements of the corrected SED are also plotted.}
            \label{fig:resid_calc}
        \end{figure*}

        Section \ref{ssec:jwst} describes the methods used to extract dereddened, intrinsic light curves from each of JWST/MIRI's four channels. However, there may exist residual corrections in the extinction law published in the \citetalias{vonFellenberg2025_extinction}, which would affect the extracted mid-IR SED. \Sg\ lends itself naturally to this quantification during flaring events assuming the flare is powered by synchrotron processes, giving a power law SED in the IR \citep[i.e.,][]{GravityCollaboration2021}. To reduce noise in this measurement, we binned each of the four channels into three frequency segments (12 total bins) using the variance-weighted mean and photometry uncertainties via standard error propagation. 

        For each frequency bin, we derived a single, multiplicative correction factor $c_i$ that is time-independent and assumed to originate only from the extinction measurement. We used a compound cost function, with a $\chi^2$ term (assuming an underlying power law model) and a Gaussian Process (GP) logarithmic likelihood regularizer term (to ensure smoothness in the corrected data). Our GP kernel is the sum of a scaled Radial Basis Function (RBF) with white-noise component \citep[\texttt{ConstantKernel * RBF + WhiteKernel} in \texttt{scikit-learn};][]{scikit-learn}, which we fit to the scaled power-law parameters. From these best-fit multiplicative factors, we estimate $0.1$ mag of residual extinction in channels 1 and 3, $0.2$ mag in channel 2, and $\leq$0.1 mag in channel 4. Recall, however, that channel 4's aperture correction (Appendix \ref{appx:apcorr}) has two ``branches.'' While the median favored the smaller value, the larger choice would impart a $\sim0.2$ mag systematic shift. Therefore, it is plausible that these two systematic shifts have canceled each other out, giving a $<0.1$ mag residual extinction error. As a consequence, we instead adopt $\sim0.2$ mag uncertainty in channel 4's extinction.
        
        We applied these residual extinction corrections to the mid-IR data for further analysis, and the residual-corrected time-dependent SED of the mid-IR flare is shown in Figure \ref{fig:timed_sed}. For comparison, the top panels of Figure \ref{fig:resid_calc}  show the uncorrected SED, and the bottom panel shows the resulting estimate, calculated as ${A}_{{V}}'\left(\nu\right) = {A}_V(\nu) - 2.5\log_{10}c_i(\nu)$, of the extinction curve in the four \texttt{SHORT} gratings. The results are consistent between the two methods within $\la$0.2~mag.
        
\begin{acknowledgments}
    We thank the anonymous referee whose comments on the submitted manuscript strengthened our results. JM is supported by an NSF Astronomy and Astrophysics Postdoctoral Fellowship under award AST-2401752. SDvF gratefully acknowledges the support of the Alexander von Humboldt Foundation through a Feodor Lynen Fellowship and thanks CITA for their hospitality and collaboration. This research was supported by the International Space Science Institute (ISSI) in Bern, through ISSI International Team project \#24-610, and we thank the ISSI team for their generous hospitality. The Submillimeter Array is a joint project between the Smithsonian Astrophysical Observatory and the Academia Sinica Institute of Astronomy and Astrophysics and is funded by the Smithsonian Institution and the Academia Sinica. We recognize that Maunakea is a culturally important site for the indigenous Hawaiian people; we are privileged to study the cosmos from its summit. This work is based [in part] on observations made with the NASA/ESA/CSA James Webb Space Telescope. The data were obtained from the Mikulski Archive for Space Telescopes at the Space Telescope Science Institute, which is operated by the Association of Universities for Research in Astronomy, Inc., under NASA contract NAS 5-03127 for JWST. These observations are associated with program \#4572. The observations are available at the Mikulski Archive for Space Telescopes (\dataset[doi:10.17909/sfb0-eq32]{\doi{10.17909/sfb0-eq32}}). Support for program \#4572 was provided by NASA through a grant from the Space Telescope Science Institute, which is operated by the Association of Universities for Research in Astronomy, Inc., under NASA contract NAS 5-03127.
    BR and BSG are supported by the Natural Sciences \& Engineering Research Council of Canada (NSERC) [funding reference number 568580], and the Canadian Space Agency (23JWGO2A01). 
    BR and AP acknowledge support by a grant from the Simons Foundation (MP-SCMPS-00001470). 
    This research was supported in part by grant NSF PHY-2309135 to the Kavli Institute for Theoretical Physics (KITP).
    NMF and DH acknowledge funding from the Natural Sciences and Engineering Research Council of Canada (NSERC) Discovery Grant program and the Canada Research Chairs (CRC) program. NMF acknowledges funding from the Fondes de Recherche Nature et Technologies (FRQNT) Doctoral research scholarship. NMF and DH acknowledge support from the Canadian New Frontiers in Research Fund (NFRF) -- Explorations program and the Trottier Space Institute at McGill. The authors acknowledge support from the Centre de recherche en astrophysique du Québec, un regroupement stratégique du FRQNT. SM is supported by a European Research Council Synergy Grant “Blackholistic” (grant 10107164).
\end{acknowledgments}

\begin{contribution}

JMM contributed to data calibration, analysis and interpretation, and manuscript writing.
SDvF contributed to analysis and interpretation, and manuscript writing.
GKK and RR contributed to developing and verifying the new SMA polarization calibration procedure.
MAG contributed to proposal and manuscript writing and scheduling the SMA observations to best-match other multiwavelength facilities. 
DH, SM, SSM, BR, GW, BSG, and AP contributed to scientific discussion and manuscript review.
TR, NMF, KH, ZS, MB, SC, MGM, GGF, JLH, and HAS, and SPW contributed to proposal and manuscript writing.


\end{contribution}
%
\facilities{SMA, JWST (MIRI/MRS)}

\software{astropy \citep{astropy:2013, astropy:2018, astropy:2022},
          CASA \citep{CASA},
          scikit-learn \citep{scikit-learn}
          }

\bibliography{main}{}
\bibliographystyle{aasjournalv7}

\end{document}